\documentclass[twocolumn,trackchanges]{aastex62} 

\usepackage{threeparttable} 
\usepackage{amsmath}
\usepackage[absolute,overlay]{textpos}
\graphicspath{{./}{figures/}}

\received{???}
\revised{???}
\accepted{???}
\shorttitle{}
\shortauthors{Han et al.}

\begin{document}
\title{The Structure, Populations and Kinematics of the Milky Way central and inner Bulge with OGLE, APOGEE and Gaia data}
\author{Xiao Han\footnote{These authors contributed equally to this work.}}
\affil{Department of Astronomy, China West Normal University, Nanchong, 637002, P.\,R.\,China}
\author[0000-0001-8459-1036]{Hai-Feng Wang\footnote{These authors contributed equally to this work.}}
\affil{Dipartimento di Fisica e Astronomia “Galileo Galilei”, Universit\'a degli Studi di Padova, Vicolo Osservatorio 3, I-35122, Padova, Italy}
\author[0000-0002-0155-9434]{Giovanni Carraro}
\affil{Dipartimento di Fisica e Astronomia “Galileo Galilei”, Universit\'a degli Studi di Padova, Vicolo Osservatorio 3, I-35122, Padova, Italy}
\author[0000-0001-6128-6274]{Mart\'in L\'opez-Corredoira}
\affil{Instituto de Astrof\'\i sica de Canarias, E-38205 La Laguna, Tenerife, Spain}
\affil{Departamento de Astrof\'\i sica, Universidad de La Laguna, E-38206 La Laguna, Tenerife, Spain}
\author[0000-0001-5082-9536]{Yuan-Sen Ting}
\affil{Department of Astronomy, The Ohio State University, Columbus, OH 43210, USA}
\affil{Center for Cosmology and AstroParticle Physics, The Ohio State University, Columbus, OH 43210, USA}
\author{Yang-Ping Luo}
\affil{Department of Astronomy, China West Normal University, Nanchong, 637002, P.\,R.\,China}
\author{Guan-Yu Wang}
\affil{Department of Astronomy, China West Normal University, Nanchong, 637002, P.\,R.\,China}

\correspondingauthor{HFW (co-first author);}
\email{haifeng.wang.astro@gmail.com;\\ };\\

\begin{abstract}
We present an analysis of the structure, kinematics, and chemo-dynamical properties of the Milky Way bulge using RR Lyrae stars from OGLE, and giant stars from APOGEE and Gaia that have distances placing them in the inner Galaxy. Firstly, using a sample of 1,879 ab-type RR Lyrae stars (RRabs) from OGLE-IV, we identified three populations: central bulge RRabs, the inner bulge RRabs, and halo or disk interlopers, based on their apocenters derived from orbital integration. Inner bulge RRabs kinematically align with the Galactic bar, while central bulge RRabs show slower rotation with lower velocity dispersion. Higher velocity dispersion stars were identified as halo/disk interlopers. Then, orbital analysis of 28,188 APOGEE Red Clump and Red Giant Branch stars revealed kinematic properties consistent with RRabs, and the chemical abundance distribution displayed a bimodal stellar density pattern, suggesting complex star evolution histories and slightly different star formation histories for the inner bulge and central bulge. The differences in the density distribution on the $|\mathrm{Z}|_{\text{max}}$-eccentricity plane for the central bulge, inner bulge, and interlopers are clearly detected. It is found that the classification of bulge stars based on orbital parameters, rather than solely on metallicity, provides a more accurate population separation. As the inner bulge, which contains the highest fraction of stars, traces the bar formed by the instability of the Galactic disk, our results support that pseudo-bulge is the primary origin of the bulge. Furthermore, fitting the observed data to both the boxy and X-shaped bulge models indicated a preference for the boxy bulge.
\end{abstract}


\section{Introduction} 

The Galactic bulge comprises multiple stellar populations with distinct structural, kinematic, and chemical characteristics, reflecting the complex dynamical processes that govern its formation and evolution \citep{2020AJ....159..270K, 2018ARA&A..56..223B}. It probably formed through the collapse of gas and/or hierarchical mergers, leading to a classical bulge, or via secular evolution driven by disk and bar instabilities, resulting in a pseudo-bulge \citep{2016PASA...33...26B}. Although both mechanisms may coexist \citep{2013ApJ...763...26O}, it remains a debated question whether both are present within the Galactic bulge \citep{2008A&A...486..177Z}.

Recent studies have utilized various stellar tracers, such as RR Lyrae (RRL) stars, Mira variables, and Red Clump (RC) stars, to investigate the Galactic bulge. RRLs, representing the oldest bulge population ($\gtrsim 12\,\mathrm{Gyr}$), serve as standard candles for determining the distance through their period-luminosity relationships. As fundamental-mode pulsators, RRab stars have well-defined pulsation properties that enable precise measurements of metallicity, reddening, and distance, making them essential for mapping the spatial structure of the bulge. \citet{2024MNRAS.530.2972S} and \citet{2020MNRAS.492.4500B} suggested that the Milky Way bar's formation can be dated by considering the oldest stars in the formed nuclear stellar disc. In the heavily obscured and densely populated bulge, bright, high-amplitude Mira variables offer valuable age estimates due to their period-age relation. Notably, Mira variables with periods around 350 days reveal a kinematic transition from nuclear stellar disc-dominated to bar-bulge-dominated dynamics. This finding supports a significant star formation burst in the nuclear stellar disc approximately 8 $\pm$ 1 Gyr ago, with possible indications of earlier, more gradual formation scenarios. \citet{2022MNRAS.517.6060S} found that younger Mira variables tend to be located at more negative Galactic longitudes, which can be attributed to the age-dependent morphology of the boxy/peanut bulge. It is important to note that observational biases in the RRL and RC samples, particularly those arising from distance errors and non-uniform sampling in longitude, significantly impact the results when using these tracers to probe the bulge \citep{2024A&A...689A.240Z}.

Understanding the structure of the Galactic bulge is essential for unraveling the formation history and evolution of the Milky Way. The bulge's morphology has been a subject of ongoing debate, with different studies suggesting various structural features. \citet{2015ApJ...811..113P} utilized OGLE-IV data to demonstrate that the bulge RRLs trace the barred structure and resemble a triaxial ellipsoidal spatial distribution, with no observed evidence of an X-shaped structure. \citet{2013ApJ...776L..19D} analyzed optical and near-infrared RRLs, revealing a spheroidal distribution with a slight central elongation, rather than a pronounced bar. In addition, a subtle bimodal density distribution was observed at high southern latitudes ($b < -5^\circ$). In the outer bulge ($8^\circ < \lvert b \rvert < 10^\circ$), \citet{2016A&A...591A.145G} supported this spheroidal distribution, but found no strong bar or X-shaped structure typical of RC stars. Extinction-corrected colour-magnitude diagrams from Baade's Window confirm that the RRL density decreases exponentially with distance from the Galactic center \citep{2019ApJ...874...30S}. \citet{2022MNRAS.509.4532S} found a peanut-shaped structure in the bulge using RRLs, with the distance between its two overdensity peaks measured at 0.7 kpc, which is significantly shorter than the 3.3 kpc obtained from near-infrared imaging and red-clump stars. In contrast with the older RRL population, younger bulge stars exhibit distinct spatial distributions. \citet{2005A&A...439..107L} derived a boxy bulge structure from 2MASS data, with axial ratios of 1:0.5:0.4 and a major axis angle of 20$^\circ$--35$^\circ$ relative to the Sun-Galactic center. The X-shaped structure in the integrated light of the bulge was revealed in WISE images presented by \citet{2016AJ....152...14N} without the need for special image processing. \citet{2016MNRAS.455.2216C} demonstrated that Mira variables with longer periods (age $\leq$ 5 Gyr) trace a clear bar, particularly in high-latitude fields. It is worth noting that Miras from OGLE survey may be incomplete for the brightest sources due to the saturation limit (at I $\sim$ 12.5) in the detector \citep{2018ApJ...865...47Q}. Using 30 million stars from Gaia DR2, \citet{2019A&A...628A..94A} estimated a bulge-bar angle of ~45$^\circ$, with a small presence of a long bar. However, \citet{2021A&A...656A.156Q} argued that this angle was overestimated and revised it to $\sim$20$^\circ$, accounting for the extinction and improved distance estimates from Gaia and APOGEE. 

The kinematics of the Galactic bulge are key to understanding its dynamical structure and the evolution of its stellar populations. \citet{2006MNRAS.370..435K} presented a proper motion (PM) mini-survey of 35 fields near the Baade window, $(l, b) = (1^\circ, -4^\circ)$, sampling roughly a $5 \times 2.5$ degrees region of the Galactic bulge. They clearly detected small gradients in PM dispersions $(\sigma_l, \sigma_b) \sim (3.0, 2.5)$ mas/yr, and in the amount of anisotropy $\left( \frac{\sigma_l}{\sigma_b} \sim 1.2 \right)$. The Bulge Radial Velocity Assay (BRAVA) data \citep{2012AJ....143...57K} confirmed cylindrical rotation in the bulge. This finding is consistent with the hypothesis of a single massive bulge formed by secular evolution, excluding a merger-made classical bulge. However, simulations by \citet{2012MNRAS.421..333S} showed that, due to the gain of angular momentum, an initially non-rotating classical bulge could transform into a fast-rotating, radially anisotropic, and triaxial object, which becomes embedded in a similarly fast-rotating, boxy bulge formed from the disk. Toward the end of this evolutionary process, the classical bulge eventually develops cylindrical rotation. Therefore, cylindrical rotation is also consistent with the dynamics of a classical bulge. \citet{2016ApJ...821L..25K} analyzed the line-of-sight (LOS) velocities of 947 RRLs, revealing hot kinematics with negligible rotation, suggesting a pressure-supported classical bulge. In contrast, N-body simulations \citep{2010ApJ...720L..72S} suggest that the bulge evolved from a bar-like pseudobulge through disk instability, with no evidence of a classical bulge formed by merger. Surveys such as ARGOS \citep{2013MNRAS.428.3660F, 2013MNRAS.430..836N} further examined bulge populations linked to disk dynamics, while the Pristine Inner Galaxy Survey (PIGS) identified potential classical bulge populations among metal-poor stars \citep{2024MNRAS.530.3391A}. 

In recent years, metallicity has been recognized as a key parameter for classifying the stellar populations within the Galactic bulge. Some observations revealed that metal-rich and metal-poor stars in the bulge exhibit disparate rotation curves \citep{2017A&A...599A..12Z,2018ApJ...858...46C,2020MNRAS.491L..11A,2021MNRAS.501.5981L}. \citet{2015ApJ...811..113P} shows that from the metallicities of the bulge RRLs, the RRLs in the inner Galaxy have three distinct old components. Using more than 15,000 RRLs from Gaia DR2, \citet{2020MNRAS.498.5629D} found that the kinematics of the Galactic bulge correlates with metallicity, with metal-rich stars exhibiting faster rotation and smaller velocity dispersions.  Analysis from the Gaia-ESO Survey (GES) revealed a metal-rich X-shaped component with bar-like kinematics and a metal-poor, extended rotating structure with higher velocity dispersion, particularly at greater distances from the Galactic plane \citep{2014A&A...569A.103R}. From the GIRAFFE Inner Bulge Survey (GIBS), \citet{2017A&A...599A..12Z} found that the metal-poor stars are more concentrated toward the Galactic center, while metal-rich stars exhibit lower velocity dispersion at a large distance from the Galactic plane ($b = -8.5^\circ$) but steeper latitude-dependent increases in dispersion. Using APOGEE and Gaia EDR3, \citet{2021A&A...656A.156Q} found bimodalities in abundance ratios such as [$\alpha$/Fe], [C/N], and [Mn/O], which suggests diverse enrichment timescales and a star formation gap between high- and low-$\alpha$ populations. Chemical maps show spatial metallicity correlations, with metal-poor, $\alpha$-rich stars concentrated toward the bulge center. \citet{2024arXiv240116711L} analyzed red giants from Gaia DR3, identifying two populations: one exhibiting bar-like kinematics and the other, Splash, characterized by a dynamically hot profile with slower rotation. The relationship between metallicity and kinematics adds complexity to the overall picture. In particular, the metal-poor component with distinct kinematic characteristics has been identified, but its origin, density distribution, and relationship to other stellar populations in the Milky Way are not well understood \citep{2016PASA...33...26B}.

It is important to note that not all stars observed toward the Galactic bulge are confined to this region, as some are halo or disk interlopers. Orbital analysis has become essential for excluding these interlopers and classifying bulge stars by their orbital parameters. Studies have shown that the extrapolated density distribution of bulge RRLs closely resembles the number density of nearby halo RRLs \citep{2017MNRAS.464L..80P}. \citet{2019MNRAS.487.3270P} conducted the first orbital analysis of 429 RRLs using six-dimensional velocity and spatial data, finding no significant differences in velocity components or orbital parameters between Oosterhoff groups. \citet{2020AJ....159..270K}, with 1,389 RRLs, classified stars according to apocentric distance, identifying bulge stars with apocenters within 3.5 kpc and halo interlopers beyond this threshold. Halo interlopers, characterized by high-velocity dispersion, affect the more metal-poor populations more significantly than the more metal-rich populations. Among the bulge stars, those located more than 0.9 kpc from the Galactic center trace the barred structure, whereas those closer than 0.9 kpc (central/classical bulge stars) do not. Similarly, \citet{2024arXiv240508990O} examined orbits of more than 4,200 RRLs and classified them by orbital time within 2.5 kpc. Their findings confirmed that metallicity is not a reliable parameter for distinguishing RRL populations in the bulge.

Despite some consensus on general results, our understanding of the structure, populations, and kinematic properties of the Galactic bulge remains incomplete. Many chemo-dynamical details remain unclear to the community:  the contribution of large building blocks and small building  blocks, the dynamics of the secular evolution, the origins of the ancient metal poor core \citep{2022ApJ...941...45R}, the X-Shape origin \citep{2012ApJ...757L...7L}, and the nature of Kraken, Heracle and Aurora populations \citep{2024MNRAS.528.3198B}, to make a few examples. This paper aims at providing some insights and more details to the Galactic bulge nature and structure using data from two surveys: OGLE-IV and APOGEE DR17. Specifically, the main point of this study is to compare the central and inner bulge stars. As one of the first papers to simultaneously compare both kinematic-dynamic properties and detailed abundances between central and inner bulge stars, it contributes to refining our understanding of the complex structure and evolution of the Galactic bulge.

This study is organized as follows.
In Section \ref{section:data}, we describe the data sources and describe the sample selection process. In Section \ref{section:Orbits}, we introduce the gravitational potential model and conduct orbital integrations for our sample. Section \ref{section:Results} presents the results of our analysis, including the kinematics, chemical properties, and the fitting of the observed density to the boxy and X-shaped bulge models. In Section \ref{section:Discussion}, we compare our orbital analysis results with those of previous studies and discuss the limitations of this work. Finally, Section \ref{section:Conclusion} summarizes the key findings and conclusions of the study.

\section{Data} 
\label{section:data}

\subsection{RRabs or RRLs of OGLE survey} 
\label{subsetction:galpys}

\begin{figure*}[!t]
  \centering
  \includegraphics[width=7in]{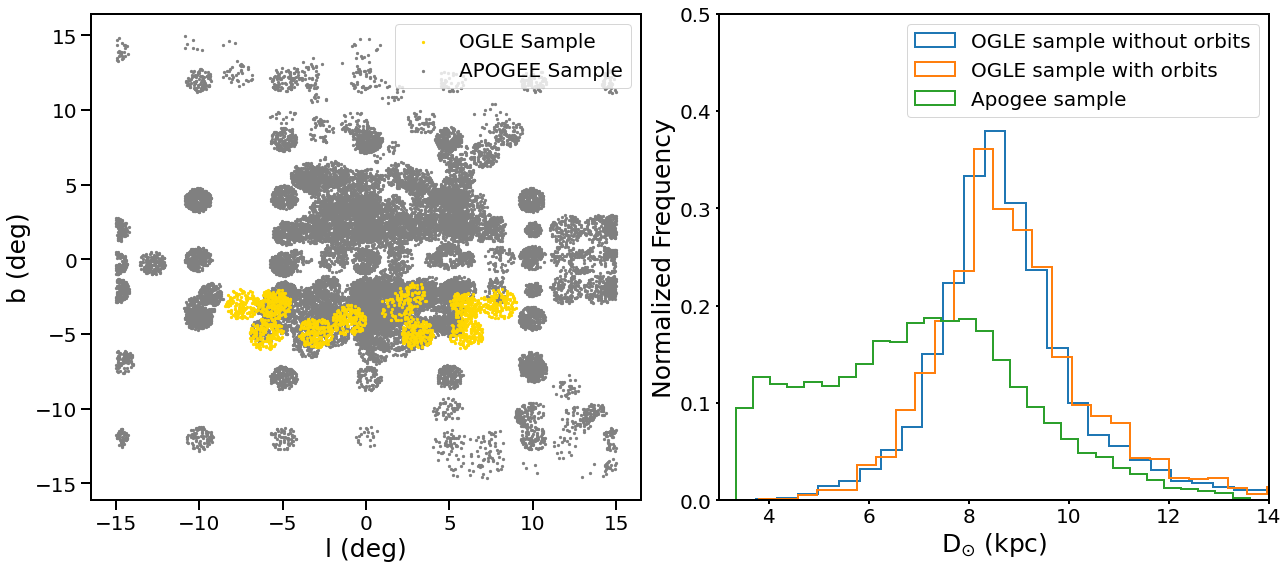}
  \caption{Left: Positions of 1,879 RRabs with orbits and 28,188 APOGEE stars in Galactic coordinates. RRab stars are represented in yellow, while APOGEE stars are shown in gray. Right: Histogram of distances to the Sun for the OGLE and APOGEE samples. The 17,817 RRLs before orbital analysis are shown in blue, the 1,879 RRLs after orbital analysis are shown in orange, and the 28,188 APOGEE stars are shown in green.}
  \label{lbscatter_disthist}
\end{figure*}

\begin{figure*}[!ht]
  \centering
  \includegraphics[width=6in]{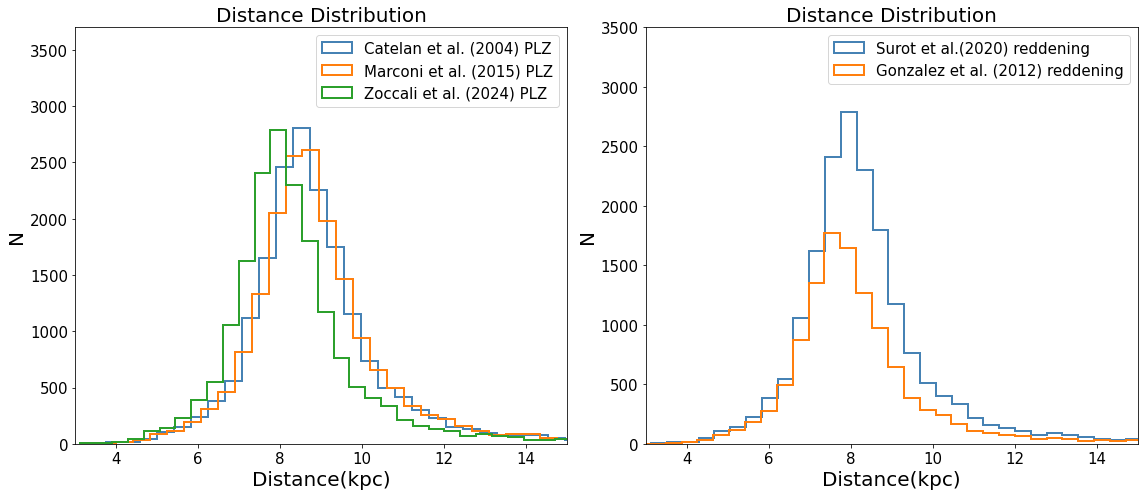}
  \caption{Histogram of distances derived using different PLZ relations (left) and different reddenings (right). Left: The blue histogram corresponds to the PLZ relation from \citet{2004ApJ...600..409C}, the orange from \citet{2015ApJ...808...50M}, and the green from \citet{2024A&A...689A.240Z}. Right: The blue histogram represents distances computed with reddening values $E(J-K_s)$ from \citet{2020A&A...644A.140S}, and the orange with those from \citet{2012A&A...543A..13G}.}
  \label{Dist_diff_PLZ_reddening}
\end{figure*}

\begin{figure}[!ht]
  \centering  
  \includegraphics[width=0.35\textwidth]{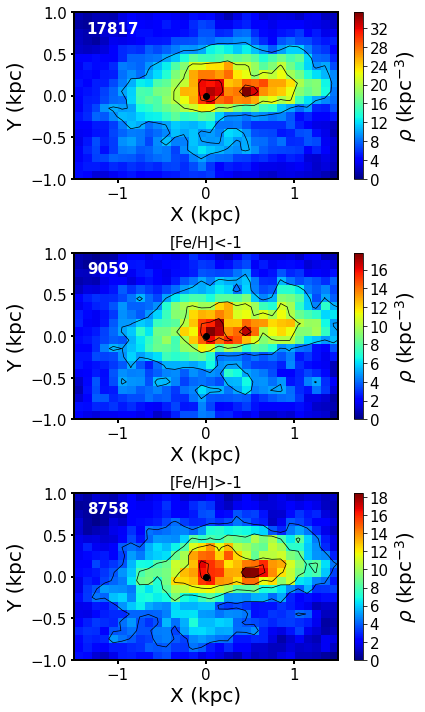}
  \caption{Top: X-Y density maps of RRab stars in a GC coordinate system for -2kpc $< \mathrm{Z} <$ 0kpc. Middle: X-Y projection of RRab density for stars with [Fe/H] $<$ -1 dex. Bottom: Same as the middle panel, but for stars with [Fe/H] $>$ -1 dex. The black dots in each panel represent the Galactic center. The star counts are labeled in the upper-left corner of each panel.}
  \label{XY_rho_FeH}
\end{figure}

\begin{figure*}[!t]
  \centering
  \includegraphics[width=7in]{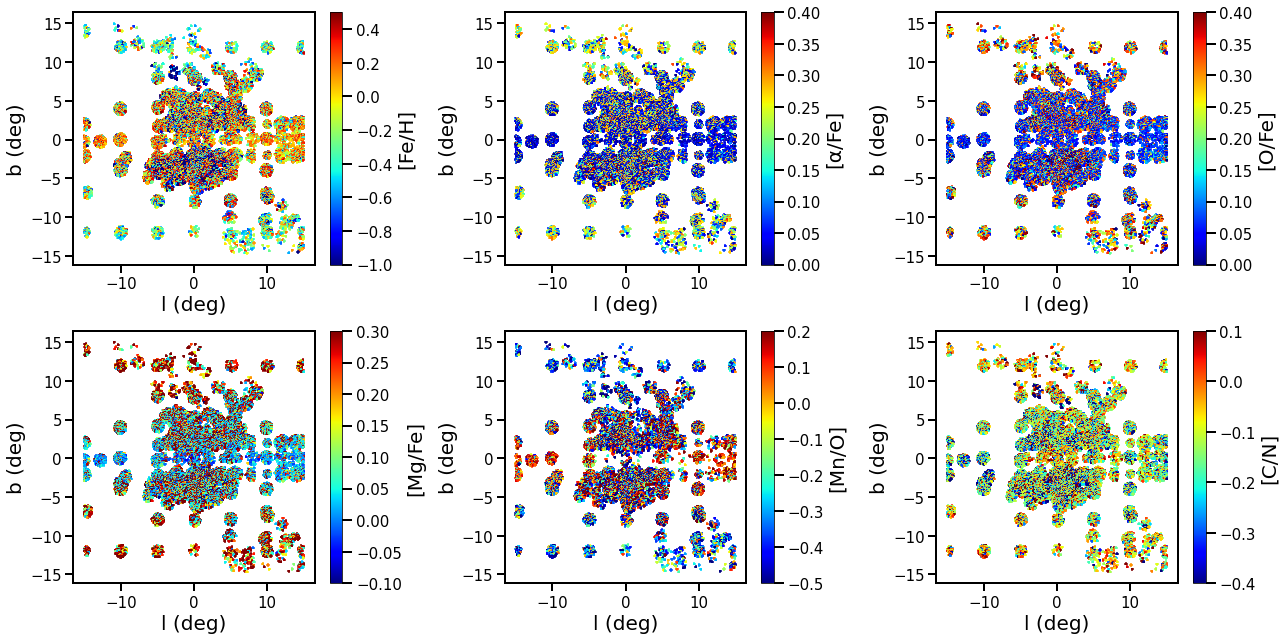}
  \caption{The distribution of sample from APOGEE in Galactic coordinates, with the colour bar representing the abundance ratios of various elements.}
  \label{llbb_element abundance}
\end{figure*}

\begin{figure*}[!ht]
  \centering  
  \includegraphics[width=6in]{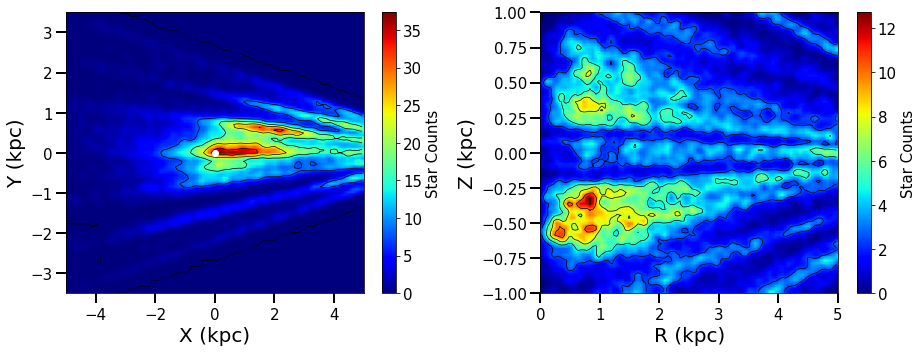}
  \caption{Density distribution of stars from APOGEE in the X$-$Y (left) and R$-$Z (right) planes in the GC coordinate system. The white dot in the left panel represents the Galactic center.}
  \label{XY_RZ_counts_APOGEE}
\end{figure*}

The original dataset employed in this study is derived from the bulge RRLs listed in the OGLE-IV catalog by \citet{2014AcA....64..177S}, focusing specifically on RRab stars which are pulsating in the fundamental mode. The catalog contains 27,258 RRabs, providing data on $I$- and $V$-band magnitudes, positions, periods and other basic parameters.

The cleaning process of RRab stars, as well as the calculation of their metallicities [Fe/H] and distances, were performed using the method described in \citet{2015ApJ...811..113P} through the period-luminosity relation of RRab. Following the paper, we independently repeated the data processing. First, we selected 22,954 RRabs from OGLE-IV and excluded 54 stars identified as true members or likely members of eight globular clusters (NGC 6441, NGC 6522, NGC 6540, NGC 6553, NGC 6558, NGC 6569, NGC 6642, and NGC 6656). Next, foreground and background RRabs were removed using a color-magnitude diagram, applying criteria \(1.1 \times (V-I) + 13.0 < I < 1.1 \times (V-I) + 16.0\) and \((V-I) > 0.3\). All this returns approximately 20,695 RRabs within the bulge. The subsequent step involved binning the \(V-I\) values and calculating the mean \(I\) magnitude within each bin. We then performed a linear fit to the binned color-magnitude data, resulting in the relation \(I = 1.15 \times (V-I) + 14.40\). The criterion of \(I < 18\) mag was applied to identify bulge members, resulting in a sample of 18,636 stars. Additionally, we applied the Galactic latitude and longitude limits of $|l| < 10^\circ$ and $-10^\circ < b <5^\circ$ to further refine the sample, ultimately obtaining 17,817 RRabs. In the cleaned data, we have known apparent magnitudes in the $V$-band and the $I$-band, as well as the periods and the combination of Fourier parameters $\phi_{31}$.

The absolute magnitudes are calculated using the Period-Luminosity-Metallicity relation (PLZ) provided by \citet{2004ApJ...600..409C}, with the following formulas:

\vspace{-\baselineskip} 
\begin{align}
    M_V &= 2.288 + 0.882 \log Z + 0.108 (\log Z)^2 \label{eq:MV} \\
    M_I &= 0.471 - 1.132 \log P + 0.205 \log Z \label{eq:MI} 
\end{align}
\vspace{-\baselineskip}

with the following conversion for metallicity:

\vspace{-\baselineskip}
\begin{equation}
    \log Z = \textnormal{[Fe/H]} - 1.765 \label{eq:logZ}
\end{equation}
\vspace{-\baselineskip}

The value of [Fe/H] is determined by the formula derived by \citet{2005AcA....55...59S}:

\begin{equation}
    \textnormal{[Fe/H]} = -3.142 - 4.902P + 0.824\phi_{31} \label{eq:FeH}
\end{equation}
\vspace{-\baselineskip}

Subsequently, the reddening in the optical system is accounted for using the definition of $E(V-I)$:

\vspace{-\baselineskip}
\begin{equation}
    E(V-I) = (V-I) - (V-I)_0 = (V-I) - (M_V - M_I) \label{eq:EVI}
\end{equation}
\vspace{-\baselineskip}

The extinction formula for the $I$-band proposed by \citet{2013ApJ...769...88N} is used:

\vspace{-\baselineskip}
\begin{equation}
    A_I = 0.7465E(V-I) + 1.3700E(J-K) \label{eq:AI}
\end{equation}
\vspace{-\baselineskip}

where the $E(J - K)$ extinction values we used are derived from \citet{2020A&A...644A.140S}. 

Finally, the distances to each RRab are obtained using the following formula:

\begin{equation}
    d = 10^{1+0.2(I_0-M_I)} = 10^{1+0.2(I-A_I-M_I)} \label{eq:d}
\end{equation}
\vspace{-\baselineskip}

The histogram of distances from the Sun for 17,817 RRabs is shown in blue in the right panel of Figure \ref{lbscatter_disthist}, with a peak located at approximately 8.3 kpc. The statistical distance error obtained through the Monte Carlo method is approximately 0.006 kpc and the systematic error is less than 0.4 kpc \citep{2015ApJ...811..113P}. Therefore, the total distance error does not exceed 6\%.

As a comparison, we discussed how the distances would change if the different absolute magnitude relations were adopted and different reddenings and extinctions were considered. Figure \ref{Dist_diff_PLZ_reddening} presents the distance histograms for RRab stars using different PLZ relations and reddenings. In the left panel, the blue histogram is based on the PLZ relation from \citet{2004ApJ...600..409C} in the I and V bands. The orange histogram uses the V-band relation from \citet{2004ApJ...600..409C} and the I-band relation from \citet{2015ApJ...808...50M}:
\[
M_I = -0.07 - 1.66 \log P + 0.17 \textnormal{[Fe/H]}
\]
The green histogram utilizes the PLZ relations from \citet{2024A&A...684A.176P} in both the I and V bands:
\[
M_V = -0.582 \log P + 0.224 \textnormal{[Fe/H]} + 0.890
\]
\[
M_I = -1.292 \log P + 0.196 \textnormal{[Fe/H]}  + 0.197
\]

In the right panel, the blue histogram corresponds to distances derived using the reddening values $E(J-K_s)$ from \citet{2020A&A...644A.140S}, while the orange histogram represents distances based on the reddening values from \citet{2012A&A...543A..13G}.

As shown in the left panel of Figure \ref{Dist_diff_PLZ_reddening}, the distances computed using the PLZ relation from \citet{2024A&A...689A.240Z} are generally smaller compared to the other two relations. From the right panel of Figure \ref{Dist_diff_PLZ_reddening}, it can be observed that the distances computed using the reddening values E(J-Ks) from \citet{2020A&A...644A.140S} are larger than those derived from \citet{2012A&A...543A..13G}. The differences in distances computed using the various PLZ relations and reddening values are not substantial and do not significantly affect the conclusions. In summary, there is no great difference in the results of our work when using the various PLZ relations and reddening values.

In this paper, we established a Cartesian coordinate system centered on the Galactic center, with the X-axis pointing toward the Sun, the Y-axis aligned with l = 90$^\circ$, and the Z-axis oriented toward the North Galactic Pole. The three-dimensional distribution of RRabs is obtained using GALPY \citep{2015ApJS..216...29B}. We restricted the sample to stars within $|\mathrm{X}| <$ 1.5 kpc, $|\mathrm{Y}| <$ 1 kpc, and -2 kpc $< \mathrm{Z} <$ 0 kpc, and divided the selected region into bins of size $\Delta$X = 0.1 kpc, $\Delta$Y = 0.1 kpc, and $\Delta$Z = 2 kpc. The number density is calculated using the following formula, where N is the number of RRabs:

\begin{equation}
    \rho = \frac{N}{\Delta X \Delta Y \Delta Z} \label{eq:rho}
\end{equation}

The top panel of Figure \ref{XY_rho_FeH} illustrates the density distribution of X$-$Y projection of the bulge RRabs. The middle and bottom panels illustrate the distribution of stars with [Fe/H] $<$ -1 and [Fe/H] $>$ -1 in the X-Y plane. Two regions of high density are observed in the top panel \citep{2013ApJ...776L..19D}. It can be observed that low-metallicity stars are mainly concentrated at X = 0, whereas high-metallicity stars are mainly concentrated at X = 0.5. This observation may help to explain the two dense regions seen in the top panel. 

\subsection{RGB and RC Stars of APOGEE survey} 
\label{subsetction:RGB and RC Stars}

We use coordinate, LOS velocities, and chemical abundance information for a range of elements provided by APOGEE DR17 \citep{2022ApJS..259...35A}. Figure \ref{llbb_element abundance} presents the distributions of the elemental abundance ratios in Galactic coordinates. The first row shows the distributions of [Fe/H], [$\alpha$/Fe], and [O/Fe], while the second row displays [Mg/Fe], [Mn/O], and [C/N]. It can be observed that as the Galactic latitude decreases, [Fe/H] and [Mn/O] increase, whereas [$\alpha$/Fe], [O/Fe], [Mg/Fe], and [C/N] are reduced.

The distances provided by \citet{2023A&A...673A.155Q}, who employed StarHorse \citep{2016A&A...585A..42S,2018MNRAS.476.2556Q,2019A&A...628A..94A} to estimate distances and extinctions for over 10 million unique stars catalogued in various public spectroscopic surveys, including 562,424 stars from APOGEE DR17. The mean uncertainty in the distances of APOGEE stars does not exceed 6\%, and the mean uncertainty in extinction is 0.178 mag \citep{2023A&A...673A.155Q}. We remark that the distance is model dependent, so the distance distribution would change with a different stellar density distribution along the line of sight.

\begin{figure*}[!ht]
  \centering
  \includegraphics[width=7in]{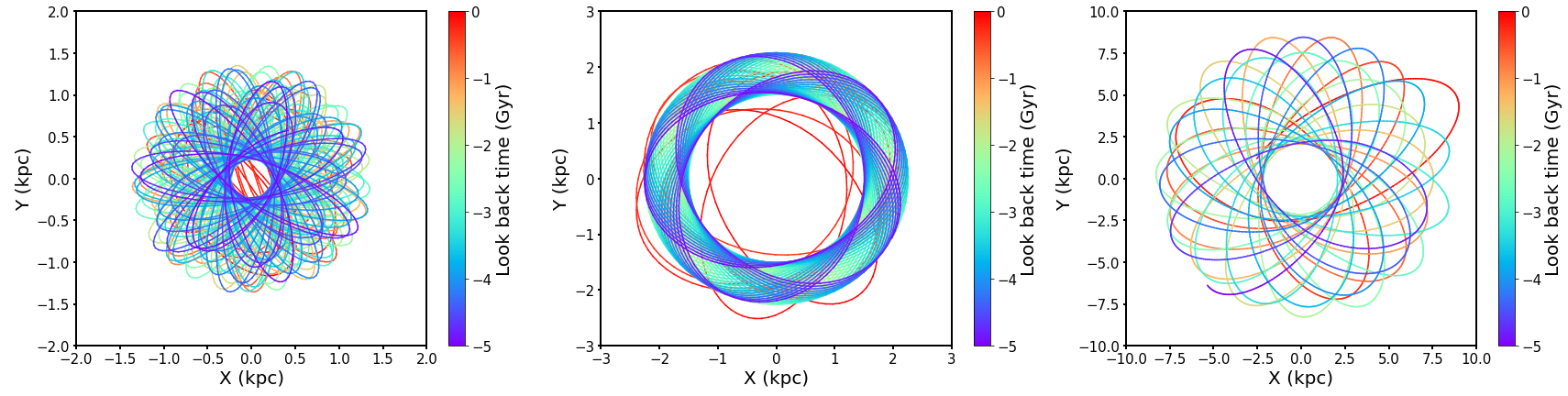}
  \caption{Top: The X-Y plane projections of the orbits of a central bulge RRL (left), an inner bulge RRL (middle), and a halo interloper (right). The color bar represents the orbital integration time.}
  \label{orbitRRL_central_inner_halo}
\end{figure*}

\begin{figure*}[!ht]
  \centering
  \includegraphics[width=6in]{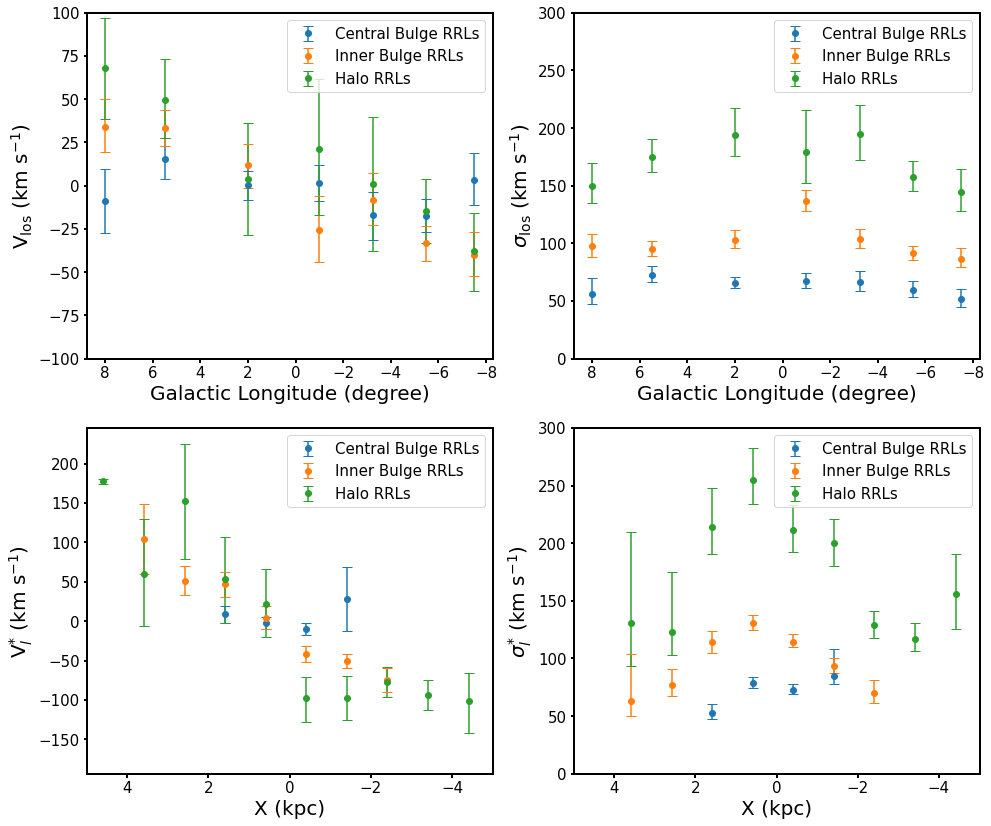}
  \caption{Top: The mean LOS velocity and velocity dispersion maps of different stellar populations as a function of Galactic longitude. Blue points represent central bulge RRabs (\( r_{\text{apo}} < 1.8 \, \text{kpc} \)), orange points represent inner bulge RRabs (\( 1.8 \, \text{kpc} \leq r_{\text{apo}} < 3.5 \, \text{kpc} \)), and green points represent halo interlopers (\( r_{\text{apo}} \geq 3.5 \, \text{kpc} \)). Bottom: The mean $v^{*}_{l}$ and velocity dispersion maps for the same stars.}
  \label{Orbit_rv_tv}
\end{figure*}

\begin{figure*}[!ht]
  \centering
  \includegraphics[width=6in]{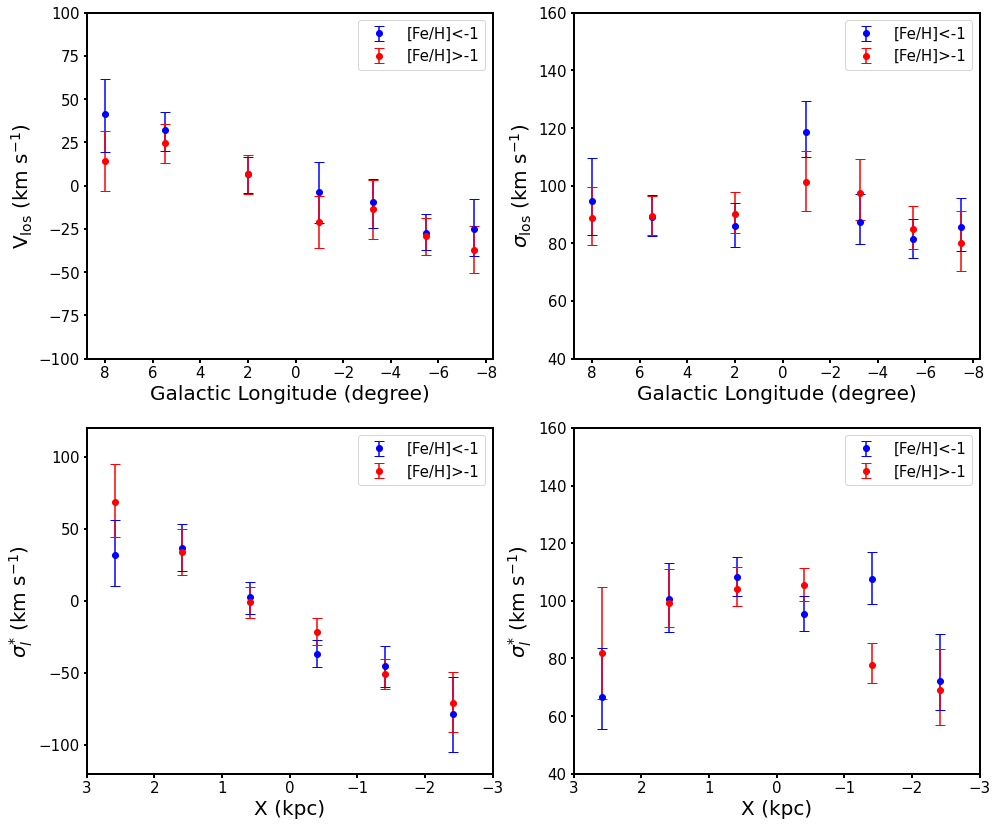}
  \caption{Top: The mean LOS velocity and velocity dispersion maps of bulge RRabs at different metallicities. The blue points represent metal-poor stars $([\text{Fe/H}] < -1$ dex), while the red points represent metal-rich stars $([\text{Fe/H}] > -1$ dex). Bottom: The mean $v^{*}_{l}$ and velocity dispersion maps of bulge RRabs at different metallicities. }
  \label{Orbit_bulge_rv_tv_FeH}
\end{figure*}

We cross-matched stars between LOS velocities from APOGEE DR17, PMs from Gaia DR3, and StarHorse distances to derive spatial velocities in Galactic Cartesian and cylindrical coordinates using GALPY \citep{2015ApJS..216...29B}. The mean uncertainty of the LOS velocities is calculated to be 0.13 km s$^{-1}$, and the mean uncertainty of the PMs is 0.16 mas yr$^{-1}$. Thanks to its high resolution and exceptionally stable platform, the instrument achieved kinematic precision that exceeds the original specifications of the survey. Specifically, the LOS velocity analysis software used in conjunction with the APOGEE instrument consistently provides LOS velocities with a precision of 0.07 km $\text{s}^{-1}$ for signal-to-noise ratios (S/N) greater than 20. Moreover, external calibration of the survey ensures accuracy at the level of 0.35 km $\text{s}^{-1}$ \citep{2017AJ....154...94M, 2015AJ....150..173N}. In order to focus on the central region of the Milky Way, the sample was restricted to stars with \(|\mathrm{X}_{\mathrm{Gal}}| < 5.0 \text{ kpc}\), \(|\mathrm{Y}_{\text{Gal}}| < 3.5 \text{ kpc}\), \(|\mathrm{Z}_{\text{Gal}}| < 1.0 \text{ kpc}\), \(|l| < 15^\circ\), and \(|b| < 15^\circ\), resulting in a total of 28,188 stars. The population in the sample is mainly composed of Red Giant Branch (RGB) and RC stars \citep{2021A&A...656A.156Q}. The spatial distribution of the samples in the (l, b) plane is displayed as gray points in the left panel of Fig. \ref{lbscatter_disthist}, while the right panel presents the histogram of the distances of the APOGEE sample from the Sun, shown in green. Figure \ref{XY_RZ_counts_APOGEE} illustrates the density distribution of the APOGEE sample, with the left panel showing the X-Y plane and the right panel showing the R-Z plane. In the X-Y plane, the sample distribution exhibits a bar-like structure with higher density around the Galactic center, while in the R-Z plane, the lack of data near the Galactic plane is likely attributed to the high extinction in this region.

\section{Orbits}
\label{section:Orbits}

Due to the overlap of bulge, disk, and halo stars near the Galactic center, it is impossible to obtain a pure sample of bulge stars based solely on distance and spatial position. Therefore, orbital analysis was employed to distinguish stars confined within the bulge from halo/disk interlopers.

We used the GALPY \citep{2015ApJS..216...29B} to define the Galactic potential model and integrate the orbits of individual stars. Our orbital integration utilized the MWPotential2014 gravitational potential model with the virial mass of M = 8.0 $\times$ 10$^{11} $M$_\odot$, which comprises the PowerSphericalPotentialwCutoff for the spherical bulge with cut-off radius of 1.9 kpc and M$_{bulge}$=0.5 $\times 10 ^{10}$ M$_\odot$, the Miyamoto-Nagai potential \citep{1975PASJ...27..533M} for the Galactic disk with scale length and height (a,b) = (3, 0.28) kpc and mass 6.8 $\times 10^{10}$ M$_\odot$, and the Navarro-Frenk-White (NFW) potential \citep{1997ApJ...490..493N} for the dark matter halo with scale radius r$_s$ = 16 kpc. Additionally, we included the KeplerPotential to represent the supermassive black hole at the Galactic center, with a mass of $4 \times 10^6$ M$_\odot$ \citep{2009ApJ...692.1075G}. We adopt the DehnenBarPotential to represent the Galactic bar, with the bar strength set to $1.1 \times 10^6$ $(\text{km/s})^2$, the bar angle set to 25 degrees, the pattern speed set to 52.25 km s$^{-1}$ kpc$^{-1}$, and the bar radius set to 3.4 kpc \citep{2000AJ....119..800D}. For all calculations, we assumed a circular velocity of $220 \, \mathrm{km \, s^{-1}}$ \citep{2012ApJ...759..131B} and a distance to the Galactic center of 8.277 kpc \citep{2022A&A...657L..12G}. We integrated the orbits backward in time for 5 Gyr.

The orbital parameters of individual stars were calculated, including the perigalactic distance (\(r_{\text{peri}}\)), apogalactic distance (\(r_{\text{apo}}\)), maximum vertical excursion from the Galactic plane (\(|Z|_{\text{max}}\)), and the eccentricity (\(e\)). We underline that different potentials will not significantly change our main physical results. The Galactic potential certainly affects the exact value for orbital parameters of the stars. However, we emphasize that while the specific values for each star’s orbital parameters may vary depending on the choice of potential, the qualitative results or patterns and the overall conclusions of our study remain unchanged. Additionally, since the uncertainties in distance and velocity are within reasonable limits, the subsequent computation and analysis of orbital parameters using GALPY are justified and acceptable.

\begin{figure*}[!ht]
  \centering
  \includegraphics[width=6in]{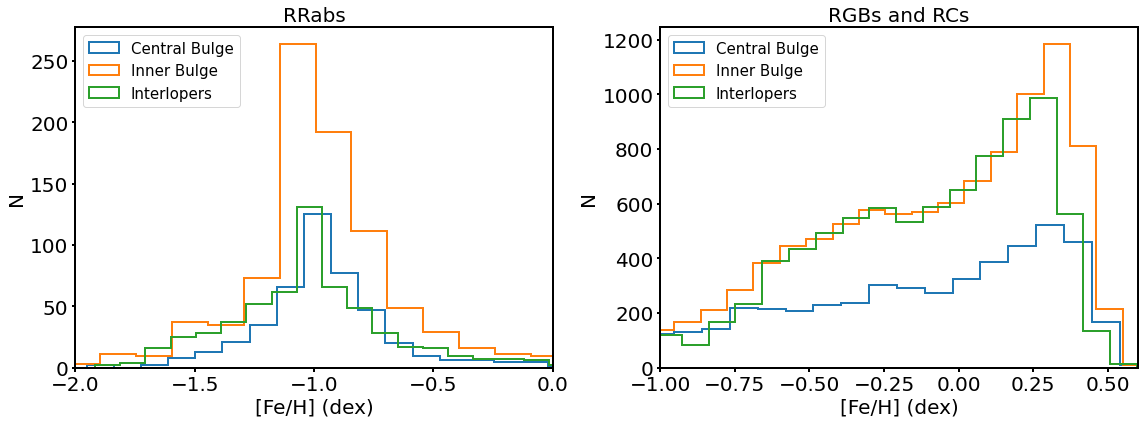}
  \caption{Left: Metallicities distribution of different stellar populations in the RRab samples from OGLE. The meanings represented by different colored points are consistent with those in Figure \ref{Orbit_rv_tv}. Right: Similar to the left panel, but for stars from APOGEE.}
  \label{FeH_Inner_Central_Halo}
\end{figure*}

\begin{figure*}[!ht]
  \centering
  \includegraphics[width=7in]{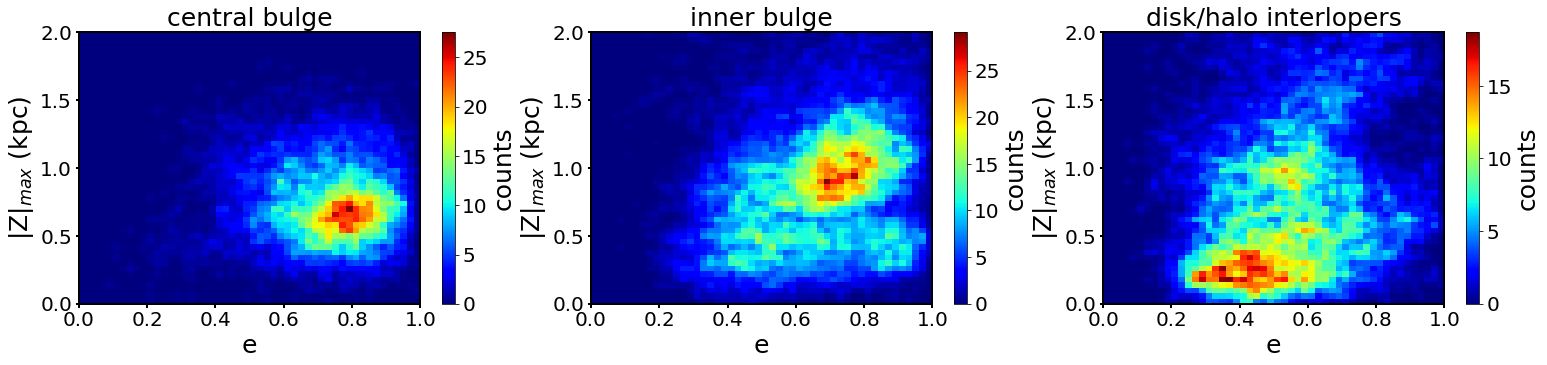}
  \caption{The distribution plots of RGBs and RCs from APOGEE survey in the $|\mathrm{Z}|_{\text{max}}$-e plane for the central bulge (left), inner bulge (right), and halo/disk interlopers (right), respectively. The color bar represents the number of stars.}
  \label{e_Zmax_counts}
\end{figure*}

\begin{figure*}[!ht]
  \centering
  \includegraphics[width=6in]{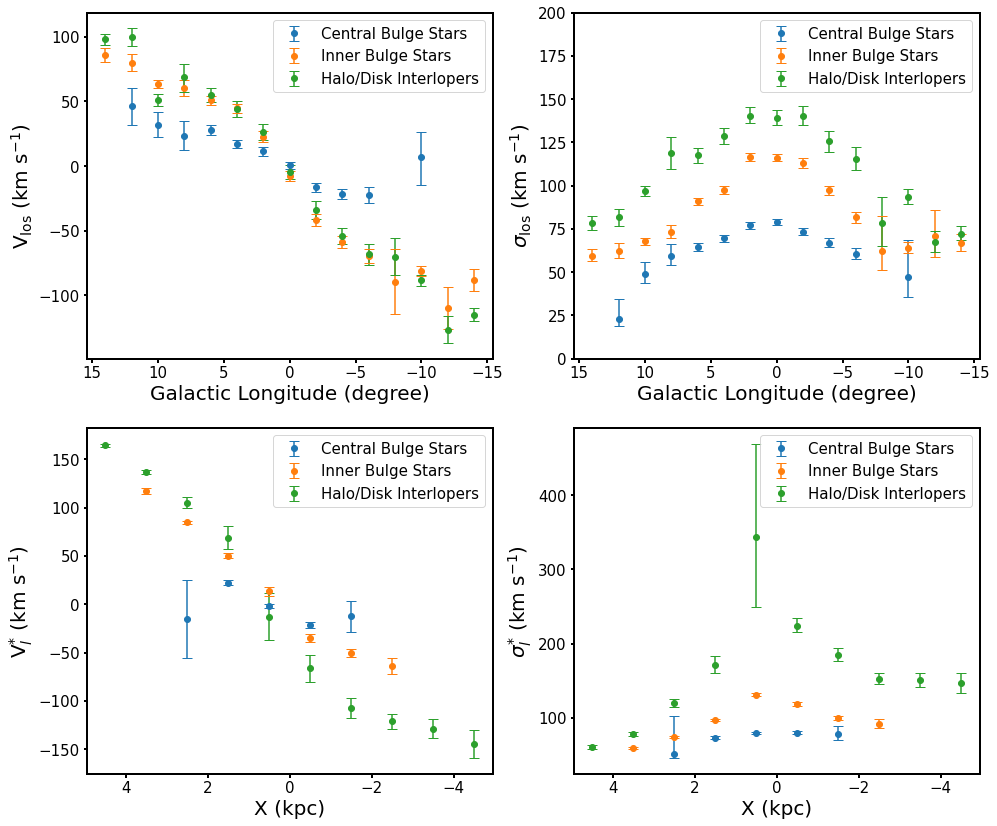}
  \caption{Top: The mean LOS velocity and velocity dispersion maps of different stellar populations as a function of Galactic longitude. Blue points represent central bulge RGBs and RCs (\( r_{\text{apo}} < 1.8 \, \text{kpc} \)), orange points represent inner bulge RGBs and RCs (\( 1.8 \, \text{kpc} \leq r_{\text{apo}} < 3.5 \, \text{kpc} \)), and green points represent halo/disk interlopers (\( r_{\text{apo}} \geq 3.5 \, \text{kpc} \)). Bottom: The mean $v^{*}_{l}$ and velocity dispersion maps for the same stars.}
  \label{lonVlos_XVll_subsetABC}
\end{figure*}

\begin{figure*}[!ht]
  \centering
  \includegraphics[width=6in]{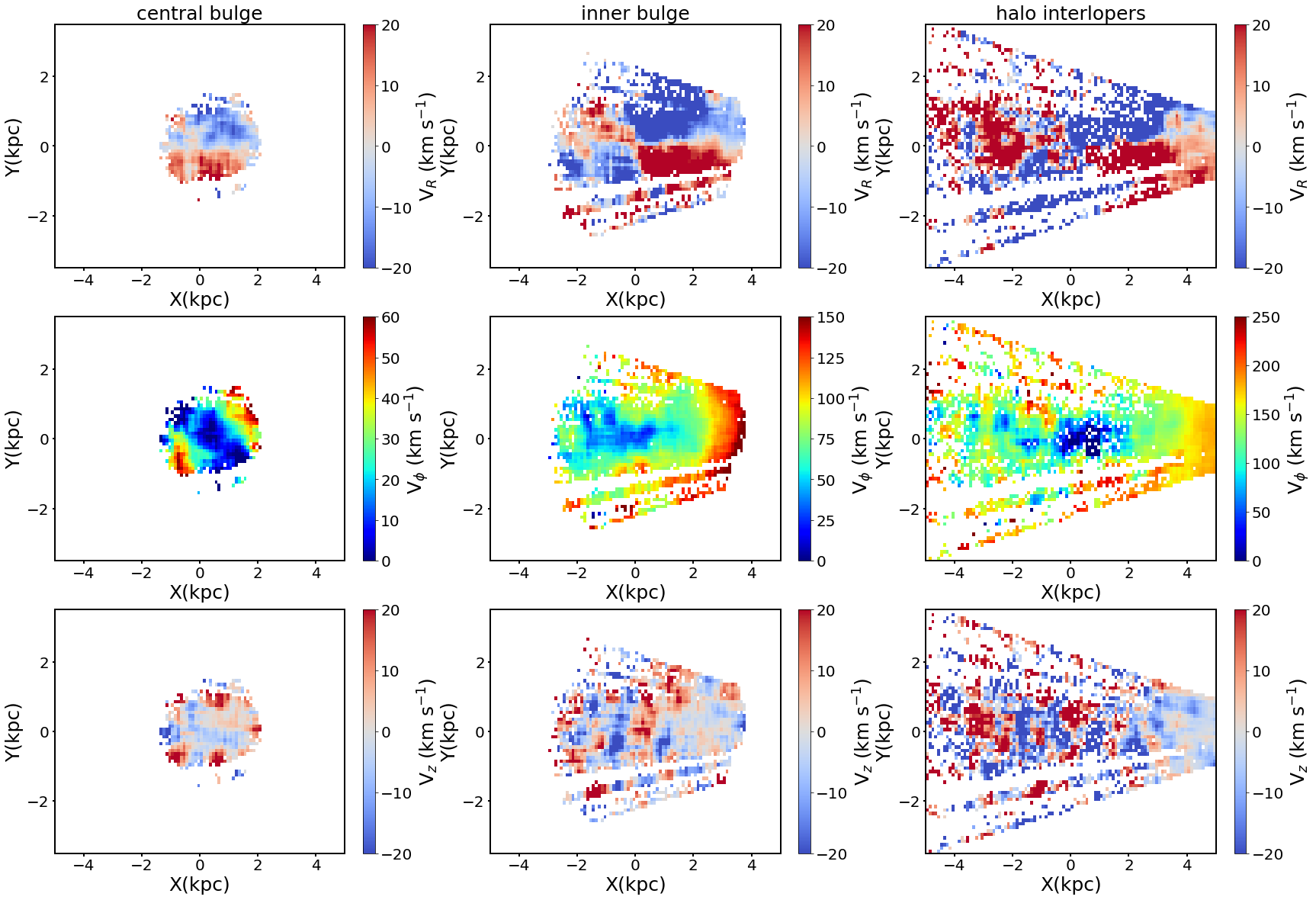}
  \caption{The velocity distribution of RGBs and RCs in the central bulge, inner bulge, and halo/disk interlopers in the X-Y plane, with $\mathrm{V_R}$ in the first row, V$_{\phi}$ in the second row, and $\mathrm{V_Z}$ in the third row.}
  \label{XY_VrVphiVz_subsetABC}
\end{figure*}

\begin{figure*}[!ht]
  \centering
  \includegraphics[width=6in]{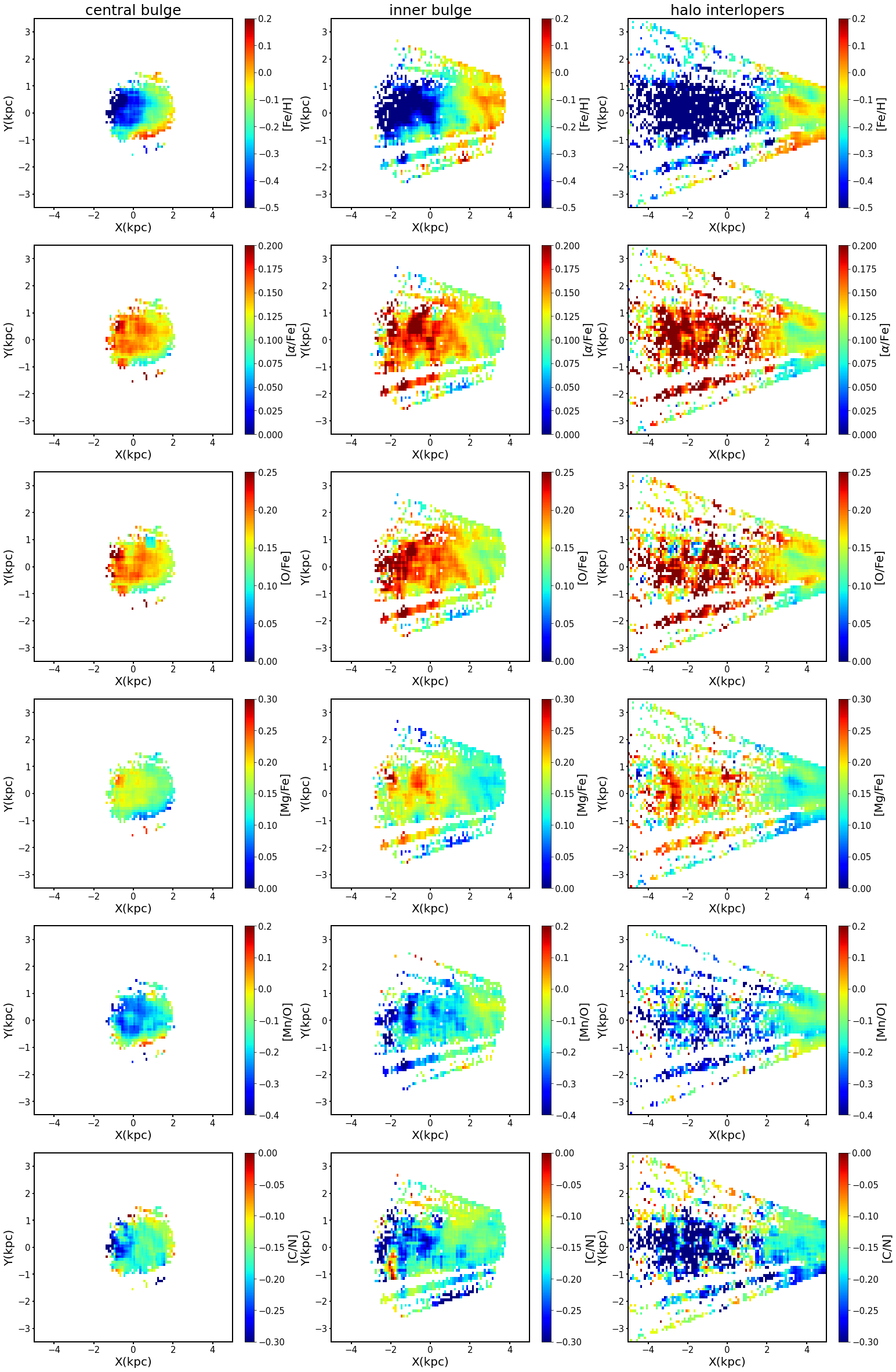}
  \caption{The chemical abundance maps of APOGEE stars in the central bulge (first column), inner bulge (second column), and halo/disk interlopers (third column) in the X$-$Y plane, with [Fe/H] in the first row, [$\alpha$/Fe] in the second row, [O/Fe] in the third row, [Mg/Fe] in the  fourth row, [Mn/O] in the fifth row and [C/N] in the last row.}
  \label{XY_metal_subsetABC}
\end{figure*}

\begin{figure*}[!ht]
  \centering
  \includegraphics[width=6.4in]{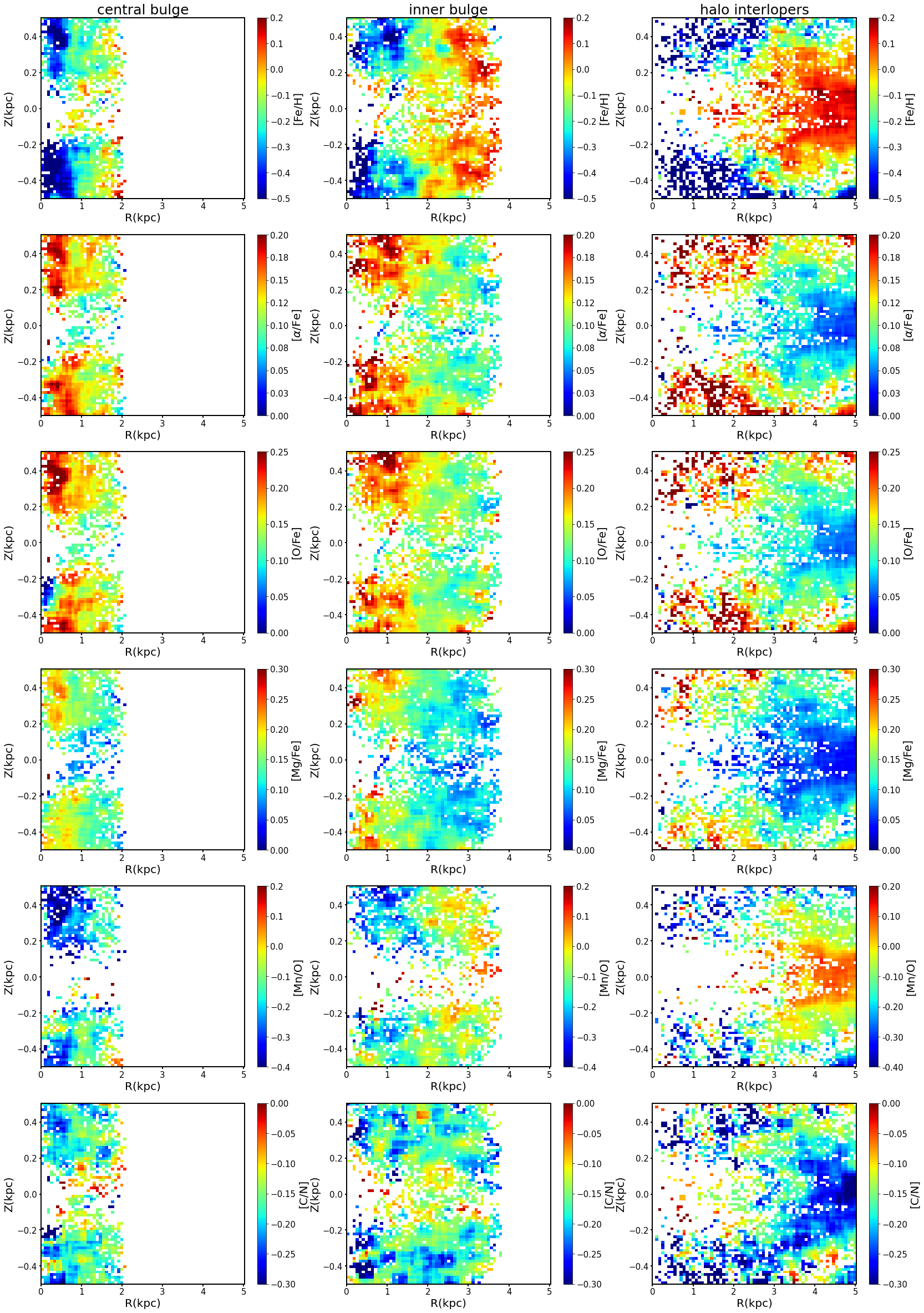}
  \caption{The chemical abundance maps of stars in the central bulge (first column), inner bulge (second column), and halo/disk interlopers (third column) in the R$-$Z plane, with [Fe/H] in the first row, [$\alpha$/Fe] in the second row, [O/Fe] in the third row, [Mg/Fe] in the  fourth row, [Mn/O] in the fifth row and [C/N] in the last row.}
  \label{RZ_metal_subsetABC}
\end{figure*}

\begin{figure*}[!ht]
  \centering
  \includegraphics[width=6.8in]{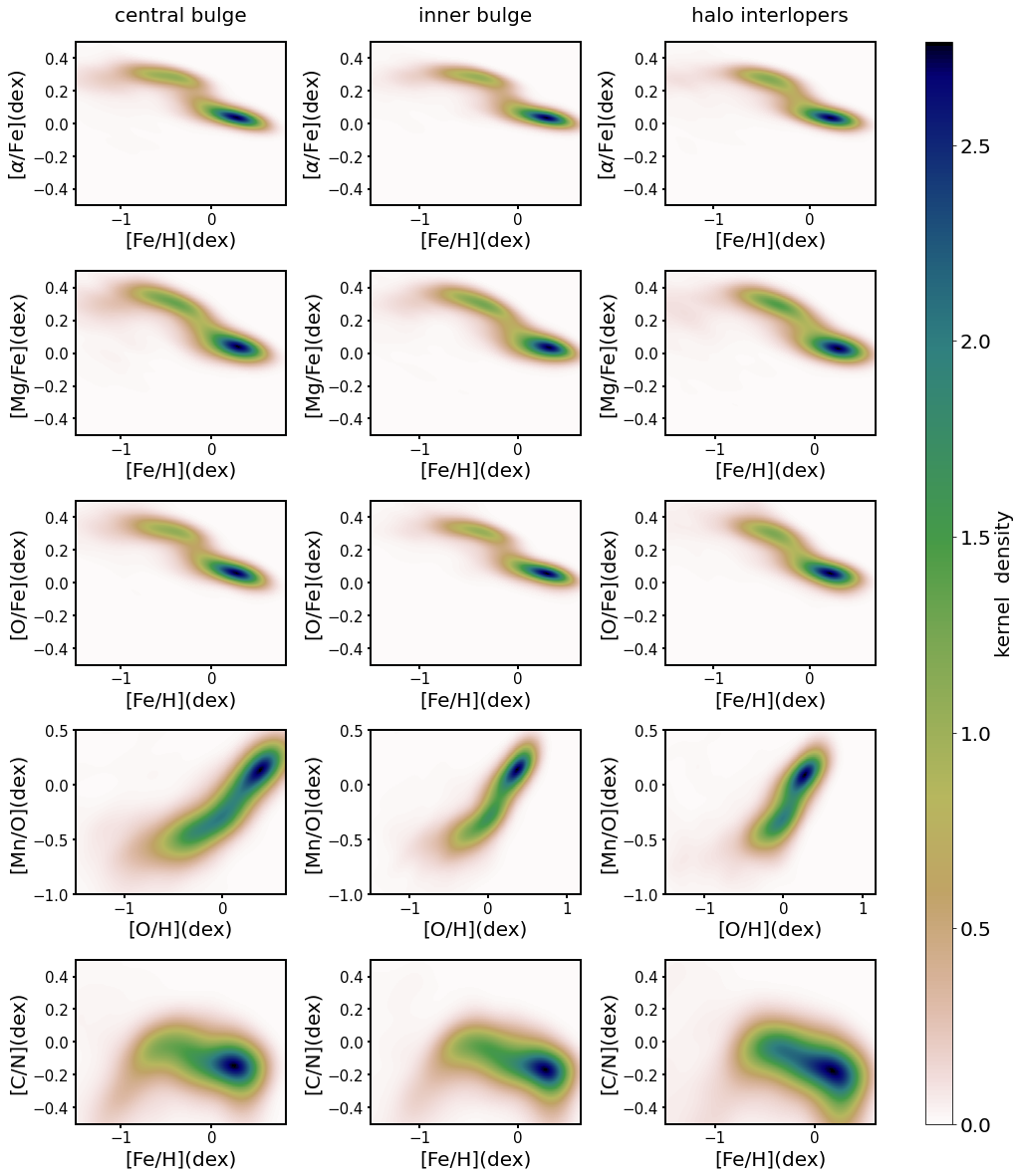}
  \caption{Two-dimensional chemical abundance plots of APOGEE RGBs and RCs in the central bulge (left), inner bulge (middle), and halo/disk interlopers (right). From top to bottom, respectively, are [$\alpha$/Fe] vs. [Fe/H], [Mg/Fe] vs. [Fe/H], [O/Fe] vs. [Fe/H], [Mn/O] vs. [O/H], and [C/N] vs. [Fe/H]. The [Mn/O] patterns suggest slightly different star formation histories for the inner bulge and central bar.}
  \label{2DMetallicity_subsetABC}
\end{figure*}

\section{Results}
\label{section:Results}
\subsection{Kinematics of RRabs from OGLE}
\label{subsection:kinematics of RRabs from OGLE}

In the study, the Local Standard of Rest (LSR) was adopted with a velocity of V$_{\mathrm{LSR}}$ = 220 km s$^{-1}$ \citep{2012ApJ...759..131B}. The distance from the Sun to the Galactic center is 8.277 kpc \citep{2022A&A...657L..12G}, and the vertical distance from the Sun to the Galactic plane is 20.8 pc \citep{2019MNRAS.482.1417B}. The solar peculiar velocity relative to LSR is (U$_{\odot}$,V$_{\odot}$,W$_{\odot}$) = (-11.1, 12.24, 7.25) km s$^{-1}$ \citep{2010MNRAS.403.1829S}.

We performed a cross-match between the RRL sample and the data from \citet{2020AJ....159..270K} and obtained LOS velocities for 2591 stars in the Heliocentric (HC) coordinate system. To avoid large dispersion in the LOS velocities, the 2,222 stars from \citet{2020AJ....159..270K} with Flag = 0 were retained. The uncertainty in LOS velocity is 10 km s$^{-1}$ \citep{2020AJ....159..270K}. The conversion from HC LOS velocity to Galactocentric (GC) LOS velocity was accomplished by 

\vspace{-\baselineskip} 
\begin{equation}
\begin{split}
    v_{\text{los},\text{GC}} &= v_{\text{los},\text{HC}} + U_{\odot}\cos l\cos b \\
    &\quad + (V_{\odot}+V_{\text{LSR}})\sin l\cos b + W_{\odot}\sin b
\end{split}
\end{equation}

Furthermore, a cross-match was conducted between the RRL sample and Gaia EDR3, identifying 13,420 stars within a matching radius of $1''$. To ensure the quality of the data, stars with a renormalized unit weight error (RUWE) greater than 1.4 were excluded. The uncertainty in proper motion is calculated to be 0.16 mas/yr. We obtain transverse velocities ($v^{*}_{l}$) using the formula $v^{*}_{l}$ = $4.74 \cdot pm \cdot dist$, where pm is the proper motion in mas yr$^{-1}$, dist is the distance in kpc, and $v^{*}_{l}$ is in km s$^{-1}$. The conversion from $v^{*}_{l,\text{HC}}$ to $v^{*}_{l,\text{GC}}$ was achieved by

\vspace{-\baselineskip}
\begin{equation}
\begin{cases}
    v_{l,\text{GC}}^{*} &= v^{*}_{l,\text{HC}}-U_{\odot}\sin l\cos b+(V_{\odot}+V_{\text{LSR}})\cos l\cos b \\
    v_{b,\text{GC}} &= v_{b,\text{HC}}+W_{\odot}\cos b \\
\end{cases}
\end{equation}
\vspace{-\baselineskip}

Following the cross-matching of the sample with $v_{los}$ and the sample with $v^{*}_{l}$ in the Galactic longitude direction, a total of 1,879 stars were identified. The coordinate transformation functions in the GALPY package \citep{2015ApJS..216...29B} were employed to derive the three-dimensional velocities of these stars. The yellow points in the left panel of Fig. \ref{lbscatter_disthist} illustrate the spatial distribution of these 1,879 RRabs in the (l, b) plane, while the orange histogram in the right panel shows their distances from the Sun.

The orbital parameters obtained in Section \ref{section:Orbits} were employed to distinguish between bulge RRabs from halo or disk interlopers. Based on classification criteria adapted from \citet{2022Univ....8..206K}, stars with \( r_{\text{apo}} < 3.5 \, \text{kpc} \) were categorized as bulge stars, with central bulge RRabs defined by \( r_{\text{apo}} < 1.8 \, \text{kpc} \) and inner bulge RRabs defined by \( 1.8 \, \text{kpc} \leq r_{\text{apo}} < 3.5 \, \text{kpc} \), while those with \( r_{\text{apo}} \geq 3.5 \, \text{kpc} \) were classified as halo interlopers. The results revealed that 24\% (451 stars) of the samples belong to the central bulge, 46\% (859 stars) belong to the inner bulge, and 30\% (569 stars) belong to the interlopers. Thus, the inner bulge contains the highest proportion of RRab stars. Typical representative stars from the central bulge, inner bulge, and halo interlopers were selected, and their orbital projections in the X-Y plane were plotted, as shown in Figure \ref{orbitRRL_central_inner_halo}. The central bulge RRL displays a compact and irregular orbit, confined to the central region. In contrast, the inner bulge RRL exhibits a nearly circular orbit with high symmetry, indicative of a low eccentricity. The halo interloper, characterized by a highly eccentric and extended path, spans a much larger region, reflecting its weaker gravitational binding to the Galactic center.

The kinematic properties of each stellar population were analyzed to provide a deeper understanding of their dynamical characteristics. The top and bottom panels of Figure \ref{Orbit_rv_tv} show the distributions of LOS velocities and $v^{*}_{l}$, respectively, for the central bulge RRabs, the inner bulge RRabs and the halo interlopers, along with their corresponding intrinsic velocity dispersion diagrams. The intrinsic dispersion of velocities can be calculated by subtracting the square of the measurement errors of the velocities from the square of the observed velocity dispersion, and subsequently taking the square root of the result. The error bars of velocity distribution are calculated using the bootstrap method to estimate the confidence intervals of the mean or standard deviation for each sample, specifically the 68\% confidence interval. It can be observed that both the average LOS velocity distribution and the $v^{*}_{l}$ distribution exhibit rotation for the inner bulge RRabs, while the central bulge RRabs show negligible net rotation. The intrinsic dispersion diagrams reveal that the central bulge exhibits the lowest velocity dispersion, followed by the inner bulge. Conversely, the high velocity dispersion is attributed to halo interlopers.

Previous studies have suggested that metal-poor and metal-rich stars in the Galactic bulge may exhibit distinct kinematic properties. To test this, we classified RRabs confined to the bulge by metallicity and analyzed the kinematics to assess whether metallicity influences velocity distributions. We divided 1310 RRLs confined to the bulge based on metallicity into metal-poor stars ($[\text{Fe/H}] < -1$, 635 stars) and metal-rich stars ($[\text{Fe/H}] > -1$, 675 stars), which are presented in Figure \ref{Orbit_bulge_rv_tv_FeH}. The top panel displays the mean LOS velocities and their intrinsic velocity dispersion, whereas the bottom panel shows the mean $v_{l}^{*}$ and corresponding intrinsic velocity dispersion. In these plots, blue points represent metal-poor stars, and red points represent metal-rich stars. It can be observed that both metal-poor and metal-rich stars exhibit similar distributions in LOS velocities, $v_{l}^{*}$, and their velocity dispersions. The left panel of Figure \ref{FeH_Inner_Central_Halo} illustrates the metallicity distribution of RRabs across different stellar populations. It is evident that the central bulge RRabs (blue), inner bulge RRabs (orange), and halo interlopers (green) exhibit no significant differences in metallicity, with peak values around [Fe/H] = -1 dex. Therefore, we conclude that metallicity is not an appropriate parameter for distinguishing different RRab populations in the Galactic bulge region.

\subsection{Kinematics and Chemistry of RGBs and RCs from APOGEE}
\label{subsection:Kinematics and Chemistry of RGBs and RCs}

The kinematic and chemical analysis of APOGEE stars is based on  research of \citet{2021A&A...656A.156Q}, with a key distinction being the classification of stars based on orbital analysis. Using the Galactic gravitational potential model from Section \ref{section:Orbits}, backward orbital integration was performed over 5 Gyr. From the classification criteria in Section \ref{subsection:kinematics of RRabs from OGLE}, approximately 21\% (5709 stars) of the sample were identified as central bulge stars, 42\% (11102 stars) as inner bulge stars, and 37\% (9783 stars) as disk/halo interlopers. Some unreliable values obtained from GALPY integration have been discarded. Consistent with the RRabs, the inner bulge stars have the highest proportion.

We analyzed stars from three populations in the $|\mathrm{Z}|_{\text{max}}$-eccentricity plane, color-coded by number of stars, as shown in Figure \ref{e_Zmax_counts}. It is observed that the central bulge stars display high eccentricities and low $|\mathrm{Z}|_{\text{max}}$, whereas the inner bulge stars are primarily concentrated at high eccentricities and high $|\mathrm{Z}|_{\text{max}}$. In contrast, the disk/halo interlopers exhibit low eccentricities and low $|\mathrm{Z}|_{\text{max}}$, consistent with the classic stellar populations of the Galactic disk.

The kinematics of the three stellar populations were analyzed and compared similar to the RRabs. Figure \ref{lonVlos_XVll_subsetABC} illustrates the LOS velocities (top) and $v_{l}^{*}$ (bottom) with intrinsic velocity dispersions for RGBs and RCs in the central bulge, inner bulge, and halo/disk interlopers. It is found that central bulge stars exhibit slower rotation and smaller velocity dispersions than inner bulge stars. The majority of stars with larger velocity dispersions are disk/halo interlopers. These results are consistent with the kinematic characteristics of RRLs. Thus, it can be inferred that the RGBs and RCs of the inner bulge trace a bar-like structure with disk-like orbits, whereas the RGBs and RCs of the central bulge do not trace the bar \citep{2020AJ....159..270K}. 

Figure \ref{XY_VrVphiVz_subsetABC} illustrates the velocity distribution of stars in the central bulge (left), inner bulge (middle), and disk/halo (right) on the X$-$Y plane. The radial velocity (V$_\mathrm{R}$), azimuthal velocity (V$_{\mathrm{\phi}}$), and vertical velocity (V$_\mathrm{Z}$) are shown from the top row to the bottom row, respectively. The inner bulge exhibits a clear quadrupole pattern in V$_\mathrm{R}$, with inward and outward motion on either side of the major axis, resembling a butterfly pattern typical of a barred bulge. In contrast, the central bulge do not exhibit this feature.

The second row of Figure \ref{XY_VrVphiVz_subsetABC} illustrates the X-Y spaces for color coding with V$_\mathrm{\phi}$. The inner bulge presents an elliptical shape, with V$_{\mathrm{\phi}}$ increasing from 0 to approximately 150 km s$^{-1}$, matching the description of a barred structure exhibiting rigid body rotation. The X-axis extends approximately 4 kpc and the Y-axis extends about 2 kpc. The bar appears to be more compact and spherical compared to the results of \citet{2015ApJS..216...29B}, confirming a similar pattern but with slightly different geometric details. In the central bulge, V$_{\mathrm{\phi}}$ ranges from 0 to 60 km s$^{-1}$, showing a relatively circular structure with lower velocities around the Galactic center. It is suggested that the central bulge is not dominated by strong rotational motion and lacks the rigid-body rotation characteristic of a bar structure. In the disk and halo interlopers plane, V$_{\mathrm{\phi}}$ is broader, reaching up to 250 km s$^{-1}$.

The third row of Figure \ref{XY_VrVphiVz_subsetABC} shows the distribution of V$_\mathrm{Z}$. In the central bulge, the V$_\mathrm{Z}$ distribution is symmetric and uniform, indicating a relatively small vertical motion. In contrast, the inner bulge exhibits an asymmetric distribution with both positive and negative values, hinting at the vertical dynamics tied to the bar structure. The V$_\mathrm{Z}$ distribution of halo/disk interlopers reveals a wave-like pattern, reflecting notable vertical oscillations within the outer disk.

The chemical composition of the bulge region is complex, as reflected in the diverse metallicity and elemental abundance patterns, which indicate a complicated history of star formation and chemical enrichment processes. The right panel of Figure \ref{FeH_Inner_Central_Halo} shows the metallicity distribution of central bulge stars (blue), inner bulge stars (orange), and halo/disk interlopers (green). It can be observed that stars confined to the bulge exhibit a peak metallicity around [Fe/H] = 0.3, while interlopers have slightly lower metallicities, peaking at [Fe/H] = 0.25. Figure \ref{XY_metal_subsetABC} and Figure \ref{RZ_metal_subsetABC} show the distribution of different chemical element abundance ratios for stars in the three populations on the X$-$Y and R$-$Z planes, respectively. From top row to bottom row, the ratios include [Fe/H], [$\alpha$/Fe], [O/Fe], [Mg/Fe], [Mn/O] and [C/N]. Both figures indicate that the elemental abundances in the bulge exhibit spatial dependence. In the X$-$Y plane, central bulge stars display metal-poor and $\alpha$-enhanced characteristics, consistent with the [O/Fe] and [Mg/Fe] ratios. In addition, lower [Mn/O] and moderate [C/N] ratios are observed. The inner bulge and the central regions of disk/halo interlopers show similar results to the central bulge. A metallicity gradient is evident with increasing distance from the Galactic center, as [Fe/H] and [Mn/O] ratios increase, while [$\alpha$/Fe], [O/Fe], and [Mg/Fe] decrease. However, the metallicity gradient of [C/N] is not clearly evident. 

The projection on the R$-$Z plane reveals that disk/halo interlopers show higher [Fe/H] and [Mn/O] values near the midplane, with these trends weakening at higher latitudes. In contrast, the trends for [$\alpha$/Fe], [Mg/Fe], [O/Fe], and [C/N] are opposite. Due to high extinction, there is a lack of data for stars in the central bulge and inner bulge at $|\mathrm{Z}| < $ 0.2 kpc, preventing the analysis of chemical abundance variations with latitude for bulge stars. For $|\mathrm{Z}| > 0.2$ kpc, consistent with the results in Figure \ref{XY_metal_subsetABC}, bulge stars exhibit chemical abundance gradients with increasing R, where [Fe/H] and [Mn/O] increase, [$\alpha$/Fe], [Mg/Fe], and [O/Fe] decrease, and the trend for [C/N] is less clear.

In Figure \ref{2DMetallicity_subsetABC}, kernel density estimation from SciPy \citep{2020NatMe..17..261V} is used to evaluate the probability density functions for stars in the central bulge, inner bulge, and disk/halo interlopers. The results are projected from top to bottom in the [$\alpha$/Fe] vs. [Fe/H], [Mg/Fe] vs. [Fe/H], [O/Fe] vs. [Fe/H], [Mn/O] vs. [O/H], and [C/N] vs. [Fe/H] planes, where Mg and O are $\alpha$-process elements, C is a light element and Mn is an iron-peak element \citep{2022MNRAS.510.2407B,2018ARA&A..56..223B}. Each subplot shows a bimodal distribution, consistent with the results observed by \citet{2021A&A...656A.156Q}, in which stars were not classified through orbital apocentric distance. The kernel density distributions of the three components are similar, suggesting complex star formation and evolutionary processes. 

In the first row of Figure \ref{2DMetallicity_subsetABC}, the bimodal distribution in the [$\alpha$/Fe] vs. [Fe/H] planes presents two distinct stellar populations in all three regions: an $\alpha$-rich group and an $\alpha$-poor group. The stellar density is dominated by metal-rich and low [$\alpha$/Fe] stars. The $\alpha$-rich stars generally formed in a period of rapid star formation and Type II supernovae (SNe), which enriched the interstellar medium (ISM) with $\alpha$-elements such as oxygen and magnesium. These stars are older and represent the early stages of the Galaxy's evolution. In contrast, the $\alpha$-poor stars probably formed after Type Ia SNe began to contribute more iron to the ISM \citep{2023A&A...673A.155Q}. Consequently, the [$\alpha$/Fe] vs. [Fe/H] plane is commonly used to distinguish between populations formed on different timescales \citep{1993A&A...275..101E,1998A&A...338..161F,2012A&A...545A..32A,2020A&A...638A..76Q,2012ceg..book.....M}. Similarly, bimodal distributions are seen in [Mg/Fe] vs. [Fe/H] and [O/Fe] vs. [Fe/H] planes, where high [Mg/Fe] and [O/Fe] values correlate with the early star formation phase and Type II SNe, while lower values reflect the later contributions from Type Ia SNe. This trend underscores the time delay between the two types of SNe and reinforces the interpretation of early versus late star formation epochs in the Milky Way \citep{2015ApJ...808..132H}.

The [Mn/O] vs. [O/H] plane provides key insights into the contributions of different SNe to the ISM. Oxygen is primarily produced by Type II SNe, while manganese production depends on both Type Ia SNe and the metallicity of the progenitor stars, making [Mn/O] a probe of the star formation history. Early stars formed under the dominant influence of Type II SNe exhibit low [Mn/O] ratios, while the increasing contribution of Type Ia SNe over time leads to a higher [Mn/O] ratio \citep{2021A&A...656A.156Q}. Additional fundamental chemical knowledge about the Milky Way can also be found in \citet{2022MNRAS.510.2407B}. This bimodal structure suggests differences in the enrichment histories of stellar populations, with older stars showing lower [Mn/O] and younger stars exhibiting higher values. The final row of Figure \ref{2DMetallicity_subsetABC}, [C/N] vs. [Fe/H] plane, reveals trends in stellar evolutionary processes. Carbon and nitrogen abundances are influenced by the dredge-up processes in stars, making this ratio a valuable tool for tracing stellar age and evolutionary status \citep{2015MNRAS.453.1855M}. In both the central and inner bulge, we observed a relatively uniform distribution with a gradual decrease in [C/N] as [Fe/H] increases. The halo/disk interlopers show a more scattered [C/N] distribution, particularly at lower [Fe/H].

In conclusion, our findings are in close agreement with those reported in previous studies. The application of orbital analysis to effectively eliminate contamination from disk/halo stars proves to be essential, as disk/halo interlopers substantially contribute to the elevated velocity dispersion observed in the Galactic bulge. For stars that are gravitationally bound within the bulge, those situated relatively far from the Galactic center display kinematic properties consistent with the bar structure. In contrast, stars situated closer to the center do not exhibit alignment with the bar dynamics.

\subsection{BP and X Shape Bulge Probability}
\label{Fits of the density}

\begin{figure*}[!htbp]
  \centering
  \includegraphics[width=6in]{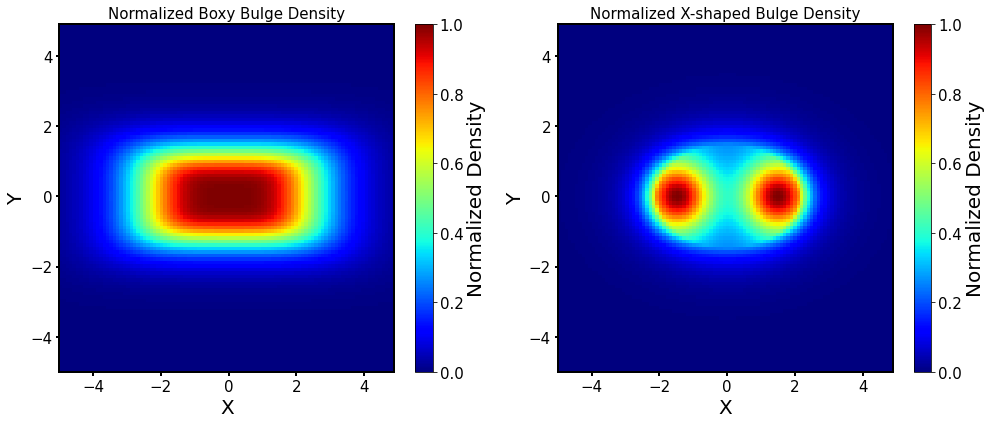}
  \caption{Density projection of the boxy bulge model (left) and the X-shaped bulge model (right) in the X-Y plane at Z=1 kpc, looks similar to \citet{2016A&A...593A..66L}.}
  \label{boxy_xshape_model}
\end{figure*}

In the previous section, we computed the density distribution of the Galactic bulge using observational data. To better understand its underlying structure, it is essential to compare these data with theoretical density models. By fitting the observed density to established models, we can evaluate which provides a more accurate representation of the bulge. We fit the stellar density to the boxy bulge model and coordinates from \citet{2005A&A...439..107L} and the X-shaped bulge model from \citet{2013MNRAS.435.1874W}, using the parametrization outlined in \citet{2017ApJ...836..218L}. The density expressions for these two models are as follows:

\begin{multline}
    \rho_{\text{Boxy}}(x, y, z) =\\
    \rho_0 \exp \left( 
    -\dfrac{\left( x^4 + \left( \dfrac{y}{0.5} \right)^4 + \left( \dfrac{z}{0.4} \right)^4 \right)^{1/4}}{740\,\text{pc}} \right)
\end{multline}

\begin{multline}
    \rho_{\text{X-shape}}(x, y, z) = \\
    \rho_0 \exp\left( -\frac{s_1}{700\,\text{pc}} \right) 
    \exp\left( -\frac{|z|}{322\,\text{pc}} \right) \times \\
    \left[ 1 + 3\exp\left( -\left( \frac{s_2}{1000\,\text{pc}} \right)^2 \right) + 
    3\exp\left( -\left( \frac{s_3}{1000\,\text{pc}} \right)^2 \right) \right]
\end{multline}

\begin{align}
    s_1 &= \max \left[ 2100\,\text{pc}, \sqrt{x^2 + \left( \frac{y}{0.7} \right)^2} \right],\notag \\
    s_2 &= \sqrt{(x - 1.5z)^2 + y^2},\notag \\
    s_3 &= \sqrt{(x + 1.5z)^2 + y^2}.\notag
\end{align}

The density maps of the boxy and X-shaped bulge models are shown in Figure \ref{boxy_xshape_model}. It can be seen that the boxy bulge model (left) exhibits a symmetric, square-like density distribution with no distinct features along the line of sight, indicating a smooth, isotropic structure. In contrast, the X-shaped bulge model (right) displays a double-peaked structure along the line of sight. The boxy-peanut shape of the bulge was a composite effect expected to appear when one considers stable orbits belonging to several families of periodic orbits \citep{2006ApJ...637..214M}. Theories of bulge formation that predict a X-shaped bulge behind the projected peanut shape appear when the bar is slowing down and the disk is thickening during bar buckling, producing the resonance to move outward and stars originally in the mid-plane are moved away from it \citep{2014MNRAS.437.1284Q}.

The least squares method was employed to fit the density \citep{2022A&A...666L..13C}. The chi-square ($\chi^2$) and reduced chi-square values ($\chi^2_{\text{r}}$) are presented below:

\begin{equation}
\chi^2 = \sum_{i=1}^{N} \frac{\left| \rho_{i,\text{model}} - \rho_{i,\text{data}} \right|^2}{\sigma_i^2}
\end{equation}

\begin{equation}
\chi^2_r = \frac{\chi^2}{N - 1}
\end{equation}

where $\sigma$ represents the density error in each bin and ($N-1$) represents the degrees of freedom. If $\rho \neq 0$, then $\sigma_i = \sqrt{\rho_{i,\text{data}} \rho_1}$, with $\rho_1 = 20 \, \text{star/kpc}^3$. If $\rho = 0$, then $\sigma_i = \rho_1$ \citep{2022A&A...666L..13C}. $\chi^2$ and $\chi^2_{\text{r}}$ are commonly used in statistical goodness-of-fit tests to evaluate the difference between model predictions and actual observational data. Only one free parameter (the amplitude) was utilised, and the remaining free parameters were not adjusted, as the sample was not intended for the precise calibration of model parameters. 

Numerical optimization methods were used on the two samples to determine the optimal value for two different models (Boxy and X-shape) by minimizing the chi-square value between the observational data and the model predictions, resulting in a better model fit. As for RRabs, the minimized reduced chi-square value for the boxy bulge model was found to be $\chi^2_{\text{r}} = 0.99$, with a probability $p = 53\%$, while for the X-shape bulge model, the minimized reduced chi-square value was $\chi^2_{\text{r}} = 1.00$, with a probability $p = 47\%$. In the case of the APOGEE sample, the minimized reduced chi-square value for the boxy bulge model was found to be $\chi^2_{\text{r}} = 0.99$, with an associated probability of $p = 55\%$. In comparison, the X-shaped bulge model yielded an minimized reduced chi-square value of $\chi^2_{\text{r}} = 1.00$, with a probability of $p = 45\%$. When the analysis was limited to the inner bulge and central bulge stars, the results remained consistent with our previous findings. For the RRab sample, the probabilities were 57\% for the boxy bulge model and 43\% for the X-shape model. For the APOGEE sample, the probabilities were 58\% for the Boxy model and 42\% for the X-shape model. These results confirm that restricting the analysis to the inner and central stars does not significantly affect the model selection. Therefore, both the RRab sample and the APOGEE sample were found to align more closely with the boxy bulge model in terms of density distribution.

\begin{figure}[!ht]
  \centering  
  \includegraphics[width=0.45 \textwidth]{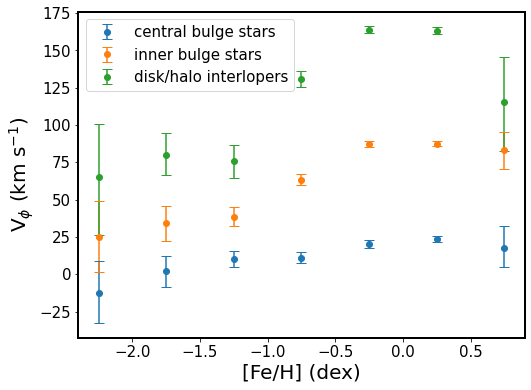}
  \caption{The variation of V$_{\phi}$ with [Fe/H] bins for the central bulge stars, inner bulge stars, and disk/halo interlopers from the APOGEE survey.}
  \label{FeH_Vphi}
\end{figure}

\section{Discussion}
\label{section:Discussion}

\begin{figure*}[!ht]
  \centering
  \includegraphics[width=6in]{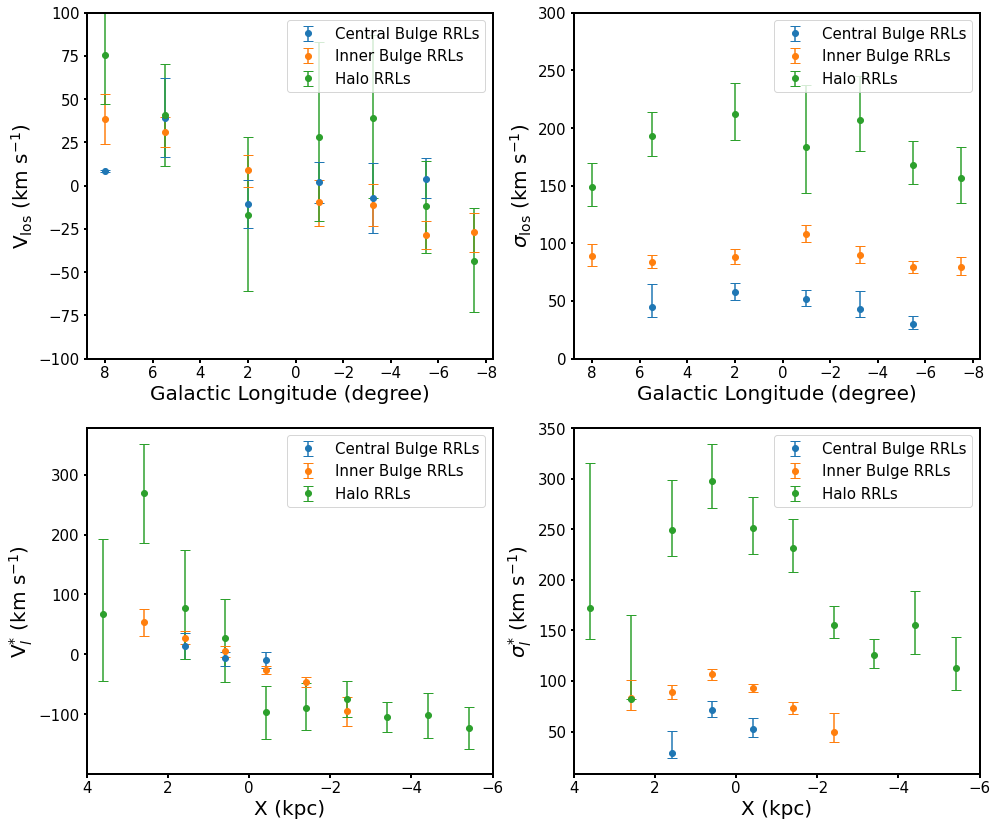}
  \caption{Similar to Figure \ref{Orbit_rv_tv}, but using the new classification criteria to classify RRabs as central bulge, inner bulge and halo interlopers.}
  \label{Orbit_rv_tv_newclassification}
\end{figure*}

\begin{figure*}[!ht]
  \centering
  \includegraphics[width=6in]{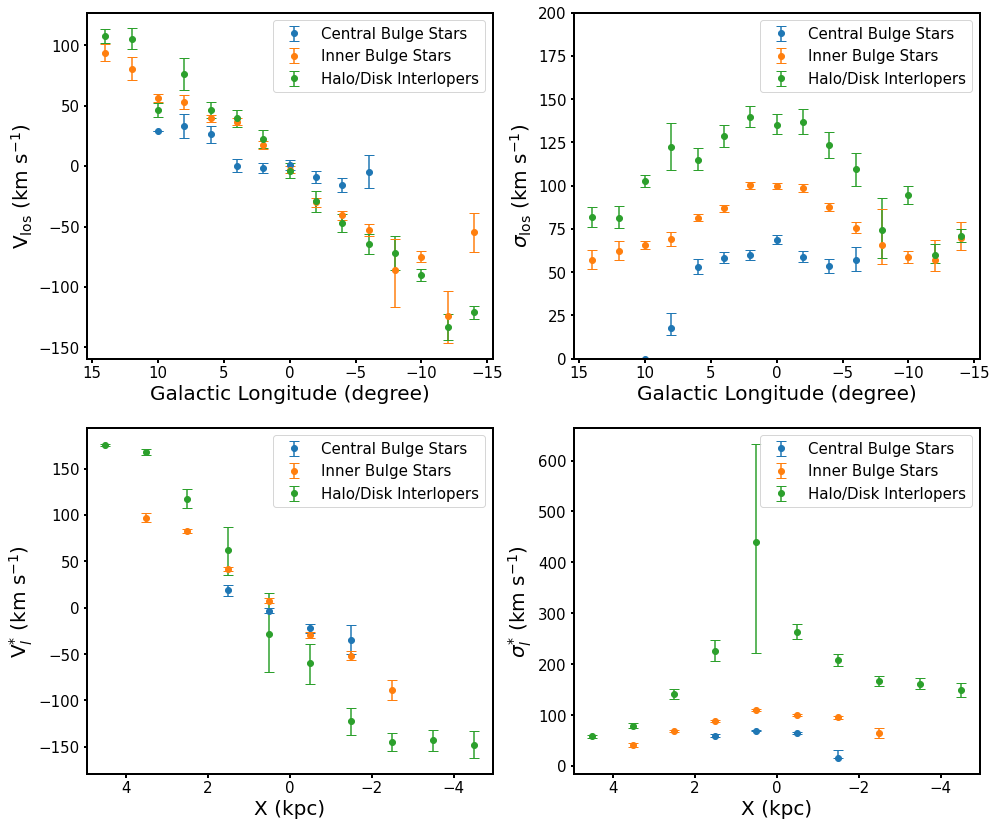}
  \caption{Similar to Figure \ref{lonVlos_XVll_subsetABC}, but using the new classification criteria to classify APOGEE stars as central bulge, inner bulge and disk/halo interlopers.}  \label{lonVlos_XVll_subsetABC_newclassification_APOGEE}
\end{figure*}

It has been reported that not all stars observed in the region of the Galactic bulge are confined to the bulge. Therefore, obtaining full spatial velocities and applying orbital analysis to remove the contamination from halo/disk interlopers is crucial. We compared the different classification criteria and results based on orbital parameters used in recent studies. 

\citet{2020AJ....159..270K} cross-matched RRLs from OGLE-IV with PMs from Gaia DR2, performing orbital analysis on the resulting sample of 1,389 stars. The orbits were integrated for approximately 5 Gyr. Stars with r$_{\mathrm{apo}} >$ 3.5 kpc was designated as halo interlopers, while those with r$_{\mathrm{apo}} <$ 3.5 kpc were classified as bulge stars. Their found 75\% of the stars confined to the bulge, slightly higher than 70\% in our sample, a discrepancy likely due to differences in the gravitational potential models used for orbital integration. Within the bulge, stars within 0.9 kpc of the Galactic center are defined as the central/classical bulge stars, while those at greater distances are classified as bar/bulge RRLs. The spatial distribution and kinematics of the bar/bulge stars align with the barred bulge structure, while the central/classical bulge stars do not follow the bar structure, which is in agreement with the conclusions of this study.

To gain further insight, \citet{2022Univ....8..206K} combined BRAVA-RR stars with Gaia eDR3 PMs and performed a similar analysis on the resulting sample of 2,628 RRLs. The boundary between the central/classical bulge and inner bulge RRLs was defined at r$_{\mathrm{apo}}$ = 1.8 kpc. Kinematically, central bulge RRLs exhibit minimal rotation and lower velocity dispersion, even at larger Galactic longitudes. In contrast, inner bulge stars show detectable rotation with a smaller amplitude but higher velocity dispersion. These findings are consistent with the results presented in this study. Spatially, neither population shows a strong association with the Galactic bar; while the less tightly bound stars exhibit a subtle alignment with the bar/bulge, the more centrally concentrated stars show little to no indication of such an alignment.

\citet{2024arXiv240508990O} combined APOGEE LOS velocities with OGLE-IV light curves and VVV PMs to calculate the orbits of a sample of 4,197 RRLs within the Galactic potential and subsequently classified the stars. Stars were classified based on their orbits relative to the bulge radius of 2.5 kpc: those remaining inside this region for at least 80\% of their orbits were identified as bulge stars, while those spending less than 30\% were classified as halo interlopers. Stars with intermediate residency (30–80\%) were left unclassified. The central bulge sample was defined as stars confined within 1 kpc for at least 80\% of their orbital period, and the inner bulge sample was defined as those within 2.5 kpc for the same fraction of time. In their sample, 57\% of stars were classified as bulge stars, 25\% as halo interlopers, and 18\% as unclassified. The proportion of stars confined to the bulge is lower than that observed in our sample, and this discrepancy may arise from differences in the adopted orbital classification criteria. The findings indicate that bulge RRLs rotate at slower speeds and have lower velocity dispersion compared to metal-rich RC stars. Their kinematics align with the low-metallicity end of the metal-poor stellar population.Further categorization of the bulge RRLs into central and inner bulge groups showed that the central bulge RRLs exhibit minimal or no rotation and lower velocity dispersion than the inner bulge RRLs, a pattern consistent with our observations. Importantly, they concluded that metal abundance is not a reliable parameter for distinguishing RRLs within the bulge region.

\citet{2021MNRAS.501.5981L} matched the spectra of 523 metal-poor stars observed with the European Southern Observatory's (ESO) Very Large Telescope (VLT) with data from Gaia DR2 to obtain three-dimensional position and velocity components, and performed orbital analysis. Consistent with \citet{2024arXiv240508990O}, they found that roughly half of the sample is associated with the bulge, while the remaining are halo interlopers. The kinematics of the metal-poor stars in the bulge were studied, revealing that their motion aligns with the characteristics of a boxy/peanut bulge and exhibits kinematic fractionation.

Based on the comparisons above, our results are in general consistent with previous studies in terms of the classification criteria and kinematic properties of bulge stars. However, there are several limitations in our analysis that warrant further consideration.  Firstly, One such limitation is that RRab stars predominantly trace old and metal-poor stellar populations, with limited representation of intermediate-age or metal-rich stars. To address this, future work focusing on increasing the sample size of metal-rich RRab stars could reveal further differences among these populations, thereby providing a more comprehensive understanding of the Galactic bulge. Secondly, while the majority of our data processing and results fall within a reasonable range, we acknowledge that there are limitations in our approach to error handling. Certain systematic uncertainties and measurement biases may still affect our analysis. Future studies incorporating more precise datasets and improved methodologies will be essential to further validate and refine our findings. Additionally, a further limitation arises from the Galactic potential model employed in this study. The MWPotential2014 model, while widely used and effective for many Galactic studies, represents an idealised approximation of the complex structure of the Milky Way. This model includes the major components of the Galaxy, such as the bulge, disk, halo and a simplified representation of the Galactic bar, but omits non-axisymmetric features such as spiral arms. The inclusion of spiral arms could significantly alter the orbital properties of stars, especially in regions where spiral arms overlap or resonate with the bar. Moreover, our analysis assumes a constant pattern speed for the Galactic bar. The bar's assumed fixed pattern speed may not reflect potential historical variations, which could influence radial migration over time; moreover, the gas disk is also ignored in this contribution. Addressing these complexities requires a more detailed, time-evolving Galactic model, which is beyond the scope of this study but is a critical direction for future work.

Furthermore, it should be noted that we only use r$_{\text{apo}}$ as a classification criterion to distinguish between the central bulge, the inner bulge, and the disk/halo interlopers. This approach possibly results in overlap and potential misclassifications among these categories, as stars with different apocentric distances are likely to occupy the same region at a given time, which may partly explain the similarity observed across some plots and panels. However, the main conclusion drawn here is based on the statistical differences in rotational velocity and dispersion, so the issue does not significantly impact the core findings of this study. As shown in Figure \ref{FeH_Vphi}, the V$_{\phi}$ distribution as a function of [Fe/H] exhibits notable differences between central bulge stars (blue points), inner bulge stars (orange points), and halo interlopers (green points) from the APOGEE survey. It can be observed that halo interlopers have the highest velocity, followed by inner bulge stars, while central bulge stars exhibit the lowest velocity. And V$_{\phi}$ increases with [Fe/H], following the same trend as shown in Figure 10 of \citet{2024MNRAS.530.3391A}.

In addition, we modified the  r$_{\text{apo}}$ classification and examined whether the results change. Referring to \citet{2024arXiv240508990O}, stars were classified based on the fraction of their orbital period spent within the bulge, which is defined by a radius of 2.5 kpc. For the RRab sample, 57\% of the stars were classified as bulge stars (1063 stars), with 89\% of them in the inner bulge (953 stars) and 11\% in the central bulge (110 stars). In contrast, 20\% of the RRL stars were classified as disk/halo interlopers (375 stars), and 23\% were unclassified (441 stars). For the APOGEE sample, 48\% of the stars were classified as bulge stars (13539 stars), with 86\% of them in the inner bulge (11649 stars) and 14\% in the central bulge (1890 stars). The remaining 29\% of APOGEE stars were classified as disk/halo interlopers (8258 stars), and 23\% were unclassified (6391 stars). Compared to the previous r$_{\text{apo}}$ classification, the proportion of stars in the inner bulge is larger than that in the central bulge , but the overall result remains unchanged. The inner bulge, which traces the bar structure formed by the instability of the Galactic disk, thus provides further support for the notion that disk secular evolution might be the primary origin of the bulge. Figure \ref{Orbit_rv_tv_newclassification} and Figure \ref{lonVlos_XVll_subsetABC_newclassification_APOGEE} present the kinematic properties of RRabs and APOGEE stars based on the revised classification scheme. The results reveal that the kinematic features remain consistent with those identified under the original classification. In particular, stars in the central bulge show lower rotational velocities and velocity dispersions compared to those in the inner bulge. Moreover, the highest velocity dispersion is still observed in the disk/halo interlopers.

\section{Conclusions}
\label{section:Conclusion}

In this study, we used the RRabs from the OGLE-IV survey in the Galactic bulge, combined with the PMs of Gaia EDR3 and the LOS velocities of \citet{2020AJ....159..270K}, to derive the three-dimensional kinematic information of 1,879 RRabs. Orbital analysis was performed using the apocentric distance to exclude halo interlopers (r$_\text{apo} \geq $ 3.5 kpc) in the Galactic bulge direction. The remaining stars were classified into two groups: inner bulge RRabs (1.8 kpc $\leq$ r$_\text{apo} <$ 3.5 kpc) and central bulge RRabs (r$_\text{apo} <$ 1.8 kpc). The inner bulge RRabs displayed kinematic consistency with the bar, while the central bulge RRabs exhibited lower rotation velocities and velocity dispersion, indicating that they do not trace the bar. The majority of stars with high velocity dispersion were found to be halo interlopers. We also observed that bulge stars with different metallicities display similar kinematic distributions, indicating that classifying bulge stars based on their orbital parameters rather than their metallicities may be more appropriate. These findings are in agreement with the results of \citet{2022Univ....8..206K}, who reported similar kinematic characteristics for RRabs in the Galactic bulge. Furthermore, compared to the work of \citet{2020AJ....159..270K} and \citet{2022Univ....8..206K}, our study employs two different types of orbital analysis and incorporates chemical parameters into the results. We have also calculated the distances, which adds further depth to our analysis.

Additionally, we used chemical abundances and LOS velocity data from APOGEE DR17, combined with PMs from Gaia DR3 and distances from StarHorse, to analyze the chemical and kinematic properties of the Galactic bulge. \citet{2021A&A...656A.156Q} conducted a detailed analysis of the kinematics, orbits, and chemical abundances of Galactic bulge stars from APOGEE DR16. They found that the stars in the central region of the bulge are characterized by low metallicity and enhanced $\alpha$-element abundances. Additionally, the chemical abundance ratios of bulge stars, such as [$\alpha$/Fe], [C/N], and [Mn/O], exhibit a bimodal distribution. We extend their findings in this study. We computed the three-dimensional kinematic distribution and orbital parameters for 28,188 stars, primarily consisting of RGB and RC stars. Similarly, we classified the sample into central bulge stars, inner bulge stars, and halo/disk interlopers, following the same criteria used for RRabs. Their kinematic characteristics were found to be consistent with the results for RRabs, indicating that inner bulge stars kinematically align with the Galactic bar, while central bulge stars exhibit slower rotation and relatively lower velocity dispersion, suggesting they are not associated with the bar structure. The disk/halo interlopers in the Galactic bulge region display a higher velocity dispersion. We examined the chemical abundances of central bulge, inner bulge and disk/halo interlopers, and derived the distribution of abundance ratios [Fe/H], [$\alpha$/Fe], [O/Fe], [Mg/Fe], [Mn/O], and [C/N] in the X$-$Y and R$-$Z planes. Our analysis also reveals a bimodal stellar density patterns of the three populations in the planes of two chemical abundance ratios.

In summary, using two different kinds of sample from different surveys, we provide additional confirmation that the bulge could be divided into the central bulge and inner bugle. The inner bulge is a bar-like population, whereas the central bulge, with lower rotation velocities and lower velocity dispersion, exhibits kinematic properties that do not align with the bar. This classification, based on orbital parameters rather than metallicity, provides a more precise understanding of the structure and dynamics of the Galactic bulge, contributing to a more detailed picture of the complex kinematic and chemical properties and star formation history of this region. Moreover, for both RRabs and APOGEE stars, the inner bulge (disk like orbits) contains the highest proportion of stars and traces the kinematic characteristics of the bar formed by the instability of the Galactic disk(e.g., \citet{2010ApJ...720L..72S}). Therefore, our results support that secular evolution is the primary origin of the bulge. Our results show that boxy/peanut (B-P) bulge population might be more dominant than X-shape bulge population.

As a future outlook, we will need more precise distance measurements in the bulge region and detailed age population data to determine the contributions of the secular evolution of the disk and spheroids. In addition, more detailed information about the impact of accretion events, such as Kraken and GSE, will be essential. 

\section*{Acknowledgements}
We acknowledge the National Key R \& D Program of China (Nos. 2021YFA1600401 and 2021YFA1600400). HFW is supported in this work by the Department of Physics and Astronomy of Padova University through the 2022 ARPE grant: {\it Rediscovering our Galaxy with machines.} MLC's research is supported by the grant PID2021-129031NB-I00 of the Spanish Ministry of Science (MICIN). Y.S.T is supported by the National Science Foundation under Grant No. 2406729. L.Y.P is supported by the National Natural Science Foundation of China (NSFC) under grant 12173028, the Sichuan Science and Technology Program (Grant No. 2020YFSY0034), the Sichuan Youth Science and Technology Innovation Research Team (Grant No. 21CXTD0038). This work has also made use of data from the European Space Agency (ESA) mission {\it Gaia} (\url{https://www.cosmos.esa.int/gaia}), processed by the {\it Gaia} Data Processing and Analysis Consortium (DPAC, \url{https://www.cosmos.esa.int/web/gaia/dpac/consortium}). Funding for the DPAC has been provided by national institutions, in particular the institutions participating in the {\it Gaia} Multilateral Agreement.

\end{document}